\documentclass{aa}  
\usepackage{txfonts}
   \newcommand{\unit}[1]{\ensuremath{\, \mathrm{#1}}}
   \newcommand{\arepo}{{\sc arepo}}          
   \newcommand{\polaris}{{\sc polaris}}          
   \newcommand{\Fig}[1]{Fig.~\ref{fig:#1}}    
   \newcommand{\Figure}[1]{Figure~\ref{fig:#1}}    

   \newcommand{\Ionetnw}{{\sf I1\_t\_nw}}

\begin{document}

   \title{Misaligned disks induced by infall}

   \subtitle{}

   \author{M. Kuffmeier
          \inst{1,2,3}\fnmsep\thanks{Marie Sk\l{}odowska-Curie Action global fellow}
          \and
          C. P. Dullemond\inst{1}
          \and
          S. Reissl\inst{1}
          \and
          F. G. Goicovic\inst{1}
          }

   \institute{Zentrum für Astronomie der Universität Heidelberg, Institut für Theoretische Astrophysik, Albert-Ueberle-Stra{\ss}e 2, 69120 Heidelberg
              \email{kueffmeier@gmail.com}
             \and
   Department of Astronomy, University of Virginia, Charlottesville, VA 22904, USA
             \and
   Max-Planck Institute for Extraterrestrial Physics, Gie{\ss}enbachstra{\ss}e 1, 85748 Garching           
          }

   \date{Received \today}

 
  \abstract
    {Arc- and tail-like structures associated with disks around Herbig stars can be a consequence of infall events occurring after the initial collapse phase of a forming star consistent with the observation of luminosity bursts. 
   An encounter event of gas with an existing star can lead to the formation of a second-generation disk significantly after the initial protostellar collapse phase. 
   Additionally, observations 
   of shadows in disks can be well described by a configuration of misaligned inner and outer disk, such that the inner disk casts a shadow on the outer disk. 
   Carrying out altogether eleven 3D hydrodynamical models with the moving mesh code \arepo, we test whether a late encounter of an existing star-disk system with a cloudlet of gas can lead to the formation of an outer disk that is misaligned with respect to the primordial inner disk. 
   Our models demonstrate that a second-generation disk with large misalignment with respect to an existing primordial disk can easily form if the infall angle is large. 
   The second-generation outer disk is more eccentric, though the asymmetric infall also triggers eccentricity of the inner disk of $e\approx 0.05$ to $0.1$.  
   Retrograde infall can lead to the formation of counter-rotating disks and enhanced accretion. 
   As the angular momentum of the inner disk is reduced, the inner disk shrinks and a gap forms between the two disks.
   The resulting misaligned disk system can survive for $\sim 100 \unit{kyr}$ or longer without aligning each other even for low primordial disk masses given an infall mass of $\sim 10^{-4} \unit{M}_{\odot}$. 
   A synthetic image for one of our models reveals shadows in the outer disk similar to the ones observed in multiple transition disks that are caused by the misaligned inner disk. 
   We conclude that late inclined infall onto a star-disk system leads to the formation of a misaligned outer disk. Infall might therefore be responsible for observations of shadows in at least some transition disks. }


   \maketitle
%

\section{Introduction}
The formation of disks is a natural consequence of the star formation process, and circumstellar disks are important as they are the birthplace of planets \citep{Keppler2018,Muller2018,Christiaens2019,Haffert2019,Pinte2018,Pinte2019,Pinte2020,Teague2018,Teague2019,Perez2020}.
Observations of Class I disks around Young Stellar Objects (YSOs) demonstrate that disks predominantly form as a by-product of a collapsing prestellar core \citep[e.g.,][]{Wolf2008,Jorgensen2009,Murillo2013,Codella2014,Ohashi2014,Maury2019}.
Using modern telescopes such as, the Atacama Large (sub-)Millimeter Array (ALMA), or the Spectro-Polarimetric High-contrast Exoplanet REsearch (SPHERE) instrument at the Very Large Telescope (VLT), it has become possible to obtain more constraints on the properties of disks. 
For instance, observations show shadows in a significant fraction of so-called transition disks that can be best explained by a misalignment between an inner and an outer disks \citep{Avenhaus2014,Marino2015,Benisty2017,Benisty2018,Casassus2018}.
Against the background of studies showing that a (sub-)stellar perturber inside \citep{Nixon2013} or outside the disk \citep{Dogan2015} can warp and break the disk, misaligned systems might be induced by a perturbing object such as a giant planet or a (sub-)stellar companion.
For instance, several hydrodynamical models show that misaligned systems can form around compact binary systems \citep{Juhasz2017,Facchini2018,Price2018}. 
Recently, \cite{Nealon2018} as well as \cite{Zhu2019} also demonstrated that a mildly misaligned planet located in the gap of a disk can lead to break-up of the disk and large misalignment of the inner disk with respect to the outer disk. 
In particular, \cite{Gonzalez2020} and \cite{Nealon2020_100453} showed that a combination of external binary and an inner planet located in the gap between inner and outer disk is a promising explanation for the specific case of HD 100453. 

However, it has also become clear that some disks, especially those around more massive Herbig stars, are associated with tail- or spiral-like structures.
One of the most prominent example of such a structure is AB Aur, which shows the presence of a large disk associated with a tail of a few 1000 AU in size, while the inner disk is misaligned \citep{NakajimaGolimowski1995,Grady1999,Fukagawa2004}. 
\footnote{Note that these large-scale structures are located beyond the spiral structures in the inner disk that are interpreted as signs of ongoing planet formation \citep{Boccaletti2020} or an inner binary companion \citep{Poblete2020}.}
Another example of such a misaligned disk around HD 100546, which hosts an extended arm structure around its disks \citep{Grady2001,Ardila2007,Walsh2017}. 
These stars have in common that they are relatively old (a few Myr) \citep{vandenAncker1998,DeWarf2003}, making it unlikely that the associated structures are remnants of the early protostellar collapse phase.
Some authors speculated that these structures might be signs of a recent flyby of an external stellar object \citep{Dai2015,Winter2018,Cuello2019}.
However, \cite{Nealon2020} find that an external stellar flyby only causes at most modest and short-lived misalignment, hence the authors rule out flyby-induced misalignment.   

\begin{figure}
    \includegraphics[width=\columnwidth]{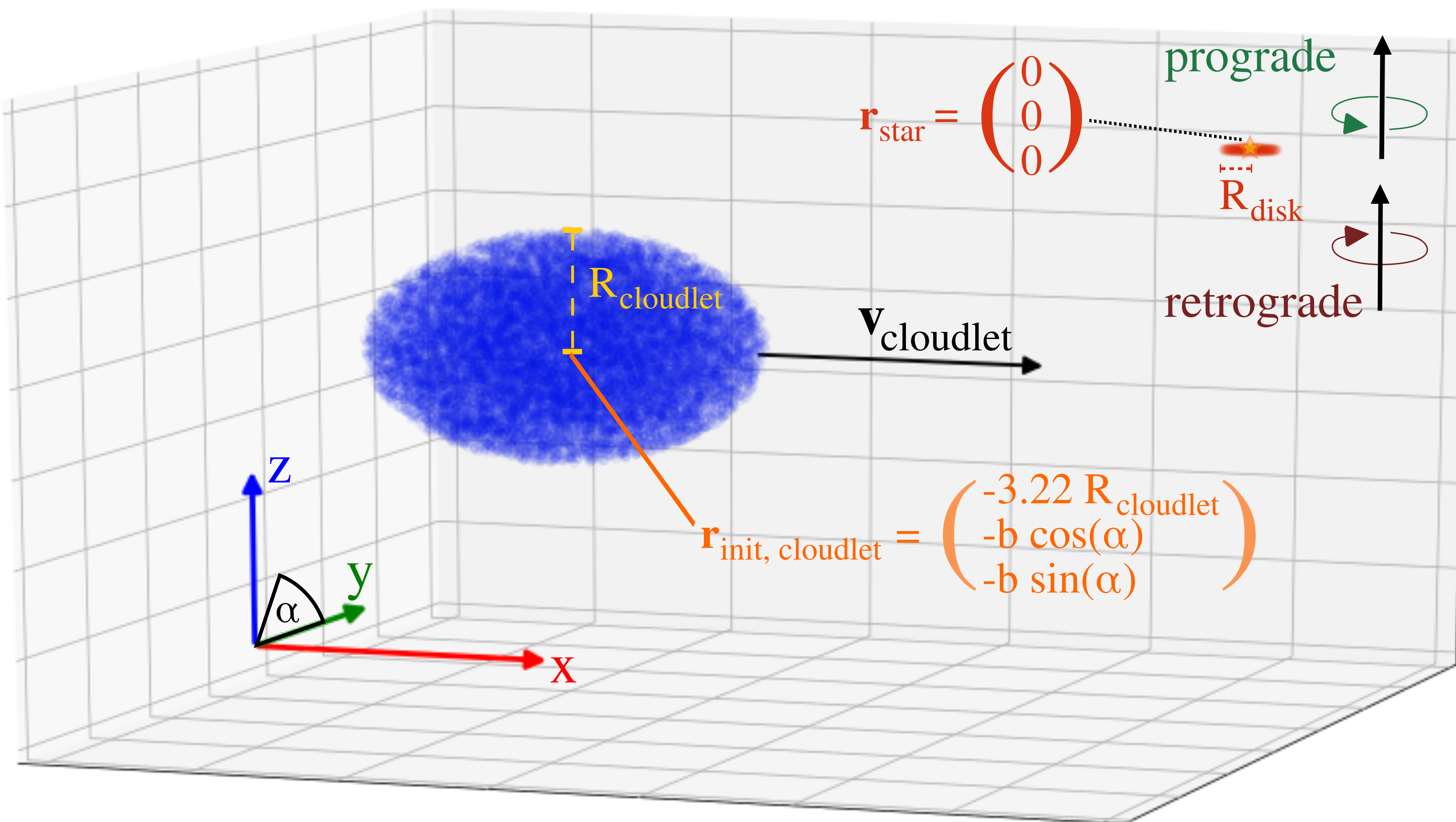} 
        \caption{Sketch of the initial setup. The initial location of the cloud varies depending on $\alpha$ and the chosen impact parameter $b$. The star is located at the origin of the coordinate system with the cirumstellar disk in the xy-plane. The disk rotation has either right-handed (corresponding to the prograde case) or left-handed orientation (corresponding to the retrograde case). Here, the illustration shows the setup for $\alpha=60^{\circ}$.} 
        \label{fig:setup-sketch}
\end{figure}

Instead, the origin of the misalignment in these systems is likely different. 
Luminosities of some evolved stars are significantly higher than stellar evolution models predict \citep{Kenyon1990}. 
Several studies showed that such enhancements in luminosity can be caused by episodic accretion events, where short-term enhancements of orders of magnitude in the accretion rate induce an increase in luminosity due to the scaling of luminosity $L\propto \dot{M}$ \citep{Vorobyov2006,Evans2009,Baraffe2009,Dunham2010}. 
In particular, luminosity bursts might be caused by late infall onto the star-disk system \citep{OffnerMcKee2011,Padoan2014,Jensen2018} that lead to disk instabilities triggering accretion bursts \citep{Kuffmeier2018}. 

Following-up on the possibility of late infall, recent numerical studies by \cite{Kuffmeier2020} (hereafter KGD20) show that disks associated with larger-scale spiral structures can form during encounter events of an existing star with surrounding gas of sufficient angular momentum \citep[see also][]{Vorobyov2020}. 
The sizes of the resulting disks crucially depend on the impact parameter of the infalling material as it is the major parameter determining the angular momentum of the gas.
This process of disk formation is different from the collapse phase leading to the formation of protoplanetary disks commonly found around young protostars.
As this process occurs at a later stage, the corresponding disks are referred to as 'second-generation disks'. 

The question is whether an infall event can not only lead to the formation of large extended structures \citep{Dullemond2019}, but also cause the formation of a system hosting a misaligned inner and outer disk as found for young forming disks in the simulations by \cite{Bate2018}.
Following-up on KGD20, we investigate the scenario in which a cloudlet of gas approaches an existing star-disk system with different inclination angles. The aim of this paper is to test whether a second-generation disk can form with different orientation around an existing inner disk.   

Section 2 describes the method used in this work. In section 3, we present the results of our numerical models. Based on the results of the models, we compare the results with observations in section 4 and discuss limitations as well as prospects of our study. In Section 5, we summarize the results and present the conclusions.   

\begin{figure*}
    \includegraphics[width=\textwidth]{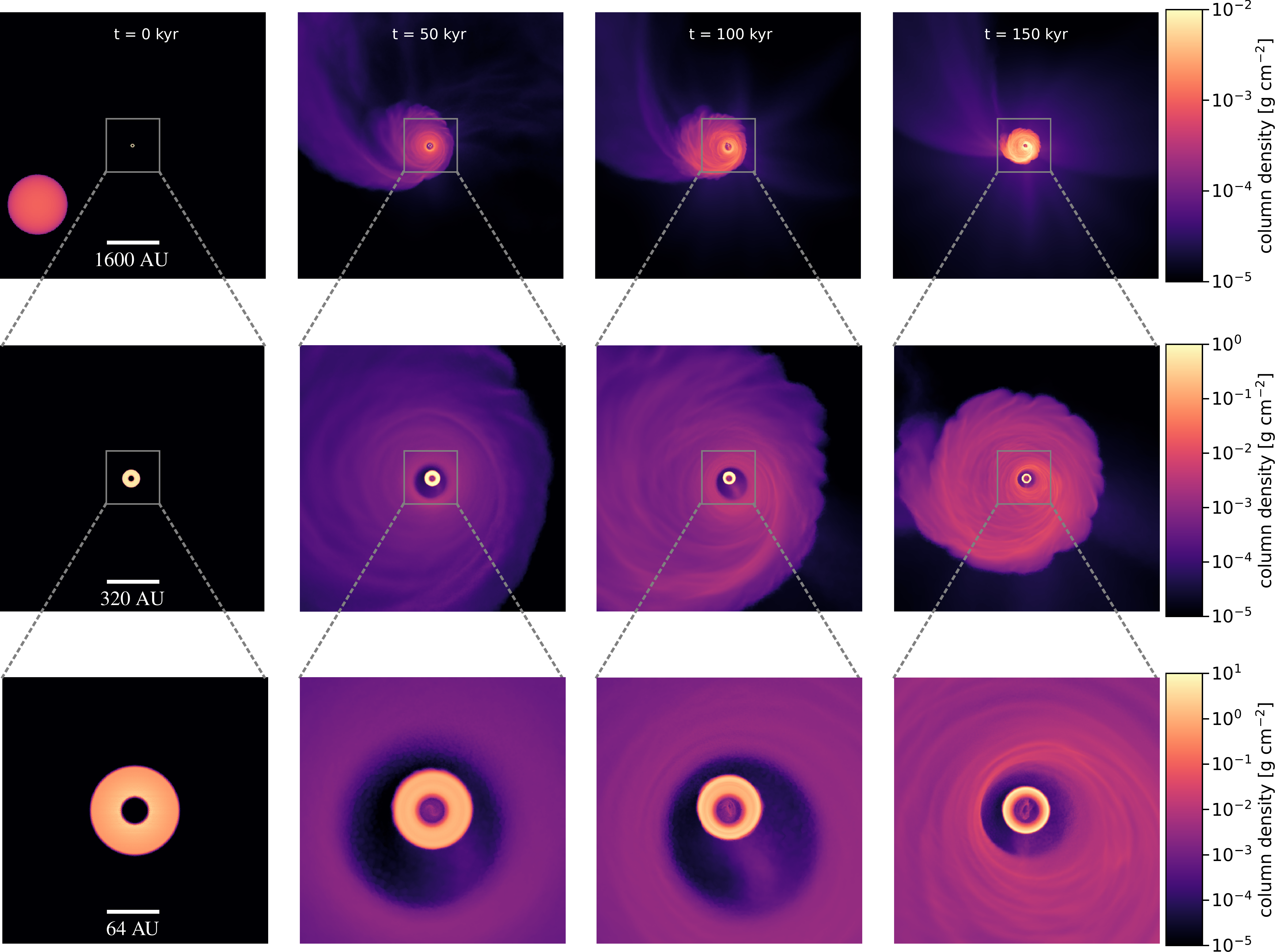} 
        \caption{Zoom-in on column densities for run 2 at $t=0$, 50, 100 and 150 kyr (from left to right) measured perpendicular to the initial plane of the inner disk with the star at the center. First row: area projection of $8000$ AU $\times\ 8000$ AU, second row: $1600$ AU $\times\ 1600$ AU, third row: $320$ AU $\times\ 320$ AU.
        }
        \label{fig:1p5_seq_000}
\end{figure*}

\begin{figure*}
    \includegraphics[width=\textwidth]{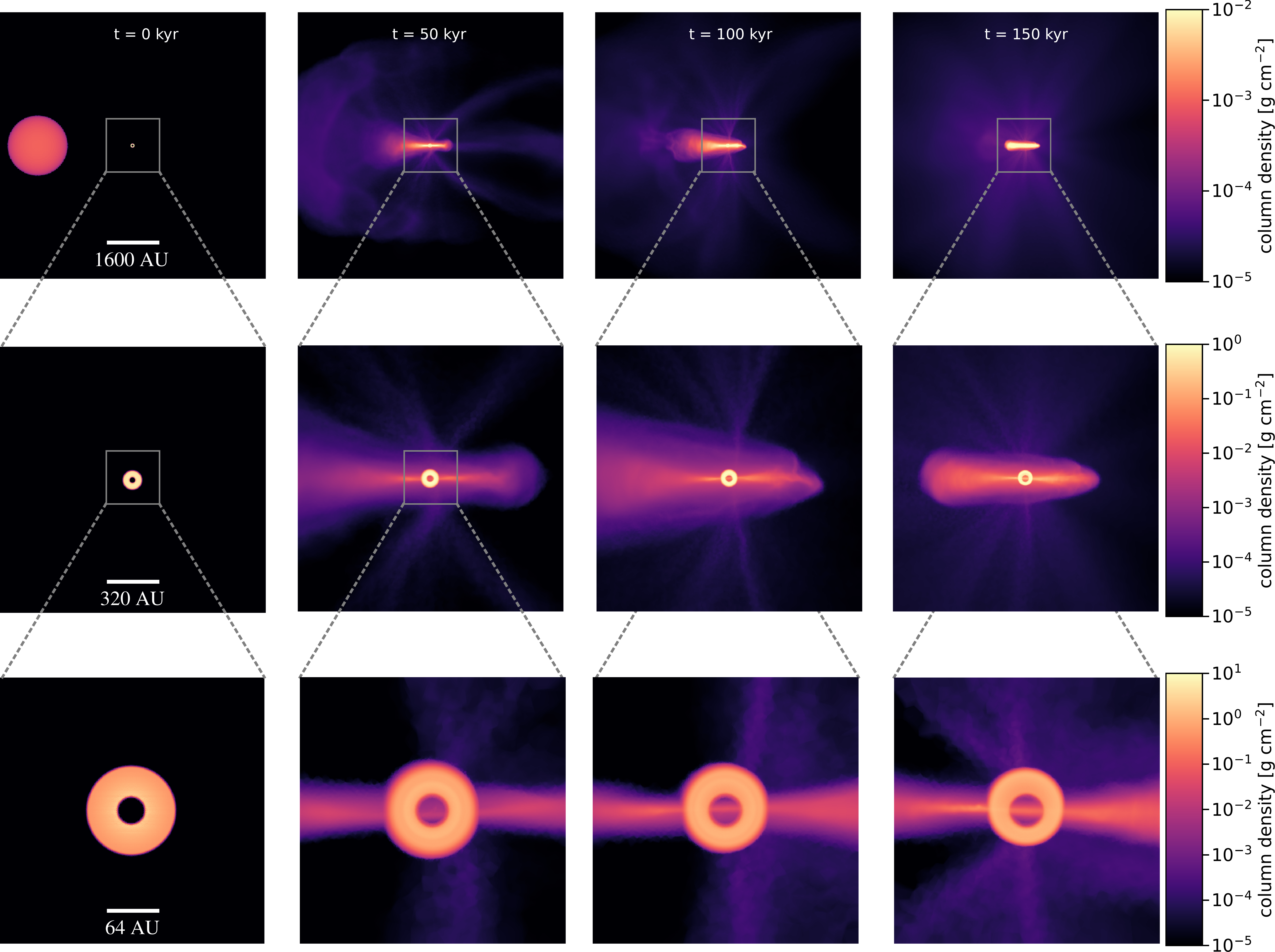} 
        \caption{Same as \Fig{1p5_seq_000}, but for run 5.}
        \label{fig:1p5_seq_090}
\end{figure*}

\section{Methods}
In this study, we analyze the effect of infall onto an already existing primordial disk. We carry out hydrodynamical simulations of a cloudlet of gas that approaches a star-disk system with different infalling angles. In a final step, we post-process one snapshot of the simulation and produce an image based on synthetic observations of scattered light and thermal emission from dust grains. We use this image to discuss our results in context with observations.

The simulations are carried out with the  moving mesh code \arepo \footnote{\url{https://arepo-code.org/}}\citep{Springel2010,Pakmor2016} solving the hydrodynamical equations and assuming the gas to be isothermal with temperature $T_{\rm gas}=10$ K. The target mass for refinement is set to $10^{24} \unit{g}$ (parameters \textsc{RefinementCriterion}\ $=1$, \textsc{DerefinementCriterion}\ $=1$, \textsc{ReferenceGasPartMass}\ $=10^{24}$ and \textsc{TargetGasMassFactor}\ $=1$ for setting the code units equivalent to \textit{cgs}-units), i.e., cells with twice/half the target mass are refined/derefined. 
The cells are allowed to have a maximum difference in volume of a factor of 5 (\textsc{MaxVolumeDiff}\ $=5$), while the minimum cell volume is set to $10^{37}\unit{cm}^{3}$ and the maximum cell volume
to $4.54\times 10^{50}$ (\textsc{MinVolume}\ $=10^{37}$, \textsc{MaxVolume}\ $=4.54\times 10^{50}$).

Analogous to the previous study presented in KGD20, we insert a star with mass $M_{*} = 2.5 M_{\odot}$ modeled as a point mass at the center of our domain and the star is fixed at its location throughout the simulations. 
We use the accretion recipe introduced in KGD20, which means that the star accretes $90 \%$ of the mass that is located within its accretion radius at every timestep. The accretion radius is set to $r_{\rm acc}=15 \unit{au}$, which is $1.5$ the gravitational softening radius of $r_{\rm grav,soft}=10 \unit{AU}$. 
The total length of the cubical box is 184 $R_{\rm cloudlet}$ and we apply periodic boundary conditions. The simulations are stopped early enough that they are unaffected by the boundary conditions.
The cloudlet approaches the star with velocity
\begin{equation}
    \mathbf{v}_{\rm cloudlet} = 
    \begin{pmatrix} 
       v_{\rm x, cloudlet} \\
       v_{\rm y, cloudlet} \\
       v_{\rm z, cloudlet} 
    \end{pmatrix} = 
    \begin{pmatrix} 
       v_{\rm i} = 10^5 \unit{cm}\unit{s}^{-1} \\
       0  \\
       0 
    \end{pmatrix},
\end{equation} 
while the background gas is at rest with respect to the central star.
The center of the cloudlet with initial radius $R_{\rm cloudlet}=887$ AU is placed at location 
\begin{equation}
    \mathbf{r}_{\rm init, cloudlet} = 
    \begin{pmatrix} 
       x_{\rm cloudlet} \\
       y_{\rm cloudlet} \\
       z_{\rm cloudlet} 
    \end{pmatrix} = 
    \begin{pmatrix} 
       -3.22\, R_{\rm cloudlet} \\
       -b \cos{\alpha}  \\
       -b \sin{\alpha} 
    \end{pmatrix},
\end{equation} 
where $b = 1774$ AU is the impact parameter and $\alpha$ is the infall angle measured with respect to the $xy$-plane of the coordinate system. $R_{\rm cloudlet}=887$ AU and $b = 1774$ AU correspond to $0.4 b_{\rm crit}$ and $0.8 b_{\rm crit}$, where $b_{\rm crit}$ is the impact parameter at which a test particle approaching a $2.5 M_{\odot}$ star with velocity $v_{\rm i}$ would be deflected by $90^{\circ}$ (for more details see section 2 in \citep{Dullemond2019}).

Adopting a slightly modified version of the description in  \cite{KlessenHennebelle2010}, the cloudlet mass is defined as 
\begin{equation}
    M_{\rm cloudlet}(R_{\rm cloudlet}) = 0.01 {\rm M}_{\odot} \left( \frac{R_{\rm cloudlet}}{5000 \unit{au}}\right)^{2.3}.
    \label{KleHen}
\end{equation}
The density of the cloudlet $\rho_{\rm cloudlet}$ is uniform and the density of the background gas is defined as $\rho_{\rm bg}=\frac{1}{800}\rho_{\rm cloudlet}$. 
To account for turbulence in the cloudlet, we set the velocity of each
gas cell according to a Gaussian random distribution using the
prescription of \cite{Dubinski1995}. The
velocities in Fourier space follow a power spectrum of 
$|\mathbf{v}_k|^2\propto k^{-4}$, with wavenumber of perturbation $k$. 
The power index is set in agreement to the velocity dispersion observed in molecular clouds \citep{Larson1981}. 
The internal velocity dispersion of the cloudlet is $10 \%$ of the initial bulk speed by multiplying each velocity by
\begin{equation}
    A_v = 0.1 \frac{v_i}{\sigma_v},
\end{equation}
where $\sigma_v$ is the velocity dispersion of field before normalization.
The setup of the infalling cloudlet is identical to run \Ionetnw\ used in KGD20 except for the initial displacement of the cloudlet if $\alpha \neq 0$. 

The major difference to the setup in KGD20 is that we insert a primordial isothermal disk ($T = 10 K$) with radius $R_{\rm disk}=50$ AU around the star at the center following the description presented in appendix A of \cite{MassetBenitez2016}. 
The disk midplane is located in the $xy$-plane of the coordinate system and has a column density profile of 
\begin{equation}
    \Sigma(r) = \Sigma_0 \left( \frac{r}{1 \unit{AU}} \right)^{-p},
\end{equation}
where $\Sigma_0$ is the amplitude of the column density and $p$ is the power index. 
Vertically (i.e., in $z$-direction), the density of the disk drops exponentially such that the density profile of the disk is defined as
\begin{equation}
    \rho_{\rm disk}(r,z) = \frac{\Sigma_0}{\sqrt{2\pi}H_0} \left( \frac{r}{1 \unit{AU}} \right)^{-(p+3/2)} e^{-z^2/(2H(r)^2)}
\end{equation}
for radii in the range of $r_{\rm in} = 20 \unit{AU} \leq r \leq R_{\rm disk} = 50 \unit{AU}$.
$H(r)$ is the scale height of the disk given by
\begin{equation}
    H(r) = \frac{c_s}{\Omega(r)},
\end{equation}
where the sound speed $c_{\rm s}$ for an isothermal gas is constant,
\begin{equation}
    c_s = \sqrt{ \frac{k_B T} {\mu m_{\rm H} } }
\end{equation} 
with Boltzmann constant $k_B$, mean molecular weight $\mu=2.3$ and hydrogen mass $m_{\rm H}$.
$\Omega$ is the orbital frequency of a Keplerian disk given by
\begin{equation}
\Omega(r) = \sqrt{\frac{GM_{*}}{r^3} }    
\end{equation}
with gravitational constant $G$, i.e., $H(r)$ scales as $H(r) \propto r^{3/2}$ for the isothermal disk considered in the simulations. $H_0$ is the scale height at $r=1 \unit{AU}$.
For $r<r_{\rm in}=20$ AU ($r_{\rm in}$ is twice the gravitational softening length of $r_{\rm grav,soft}=10 \unit{AU}$), the disk profile is tapered off such that the density drops according to
\begin{equation}
    \rho_{\rm disk,in}(r,z) = \rho_{\rm bg} + \frac{\rho_{\rm disk}(r=r_{\rm in},z) - \rho_{\rm bg}}{1 + e^{-k \left(\frac{r - r_{\rm in}}{1 \unit{AU}} + \frac{d_{\rm dec}}{1 \unit{AU}} \right)}}, 
\end{equation}
where $k=3.5$ and $d_{\rm dec}=2 \unit{AU}$ determining the sharpness of the density drop. 
For these values of $k$ and $d_{\rm dec}$, the density at $r=r_{\rm in}-\epsilon=20 \unit{AU}-\epsilon$ (where $\epsilon$ is an arbitrarily small number) is $\approx 99.9 \%$ of $\rho_{\rm disk,in}(r=20 \unit{au})$.
At radius $r=r_{\rm in}-d_{\rm dec}=18 \unit{AU}$, the density drops to $\approx 50 \%$ of its maximum at $r=20 \unit{AU}$ plus the negligible contribution of the background density, i.e., $\rho_{\rm disk}(18 \unit{AU},z) = 0.5 \times (\rho_{\rm disk}(20 \unit{AU},z) + \rho_{\rm bg})$. 
To avoid spurious effects from cells in the vicinity of the star, we force high refinement of the low-density gas located in the inner region within $r_{\rm in}$ by applying the criterion described in appendix B of KGD20. 
Similarly to the treatment for the region inside the disk, the density is tapered off exterior to $r>R_{\rm disk}=50 \unit{AU}$ according to
\begin{equation}
    \rho_{\rm disk,ext}(r,z) = \rho_{\rm bg} + \frac{\rho_{\rm disk}(r,z) - \rho_{\rm bg}}{1 + e^{k \left(\frac{r - R_{\rm disk}}{1 \unit{AU}} - \frac{d_{\rm dec}}{1 \unit{AU}} \right)}}, 
\end{equation}
such that the density at $R_{\rm disk}+\epsilon$ is $\approx 99.9 \%$ of $\rho_{\rm disk}(r=R_{\rm disk},z)$ and at $r=R_{\rm disk} + d_{\rm dec} = 52 \unit{AU}$, it is $\approx 50 \%$ of the density that the disk would have at this radius without the introduction of an outer edge.
We use logistic functions for the tapering to ensure that the density profile asymptotically approaches both the values at the inner/outer edge of the disk as well as the minimum density of $\rho_{\rm bg}$.
To test the effect of the infalling angle of the cloudlet onto the disk, we vary the initial position of the cloudlet angle with the angle $\alpha$. 
The tested angles are $\alpha = 0^{\circ}$, $35^{\circ}$, $60^{\circ}$ and $90^{\circ}$.  
The primordial disk is on all cases aligned with the $xy$-plane. 

In this study, we define $\Sigma_0=170 \unit{g}\unit{cm}^{-2}$ for the main runs, which is $0.1$ times the column density at 1 AU of the minimum mass solar nebula and we perform simulations using $p=1.5$.
The rotation axis of the inner disk is parallel or anti-parallel to the positive $z$-axis of the coordinate system.
A parallel/anti-parallel configuration corresponds to prograde/retrograde infall.
To test the influence of the inner disk mass, we also perform simulations with very low inner disk mass of $\Sigma_0=17 \unit{g}\unit{cm}^{-2}$ for an infalling angle of $\alpha=90^{\circ}$, and without inclination $\alpha=0^{\circ}$ for prograde or retrograde infall.
All simulations are carried out for $t=150 \unit{kyr}$ of evolution.
Moreover, we carry out a comparison run for with $\alpha=90^{\circ}$ inclination for a shorter time sequence of $t=100$ kyr using $\Sigma_0=170 \unit{g}\unit{cm}^{-2}$, but $p=1$.    
The chosen parameters for the altogether eleven runs are summarized in table \ref{runover}. 
\begin{table}[th]
\centering
\begin{tabular}{ccccccc}
     Run   & $R_{\rm disk}$ & $\alpha$ & $\Sigma_0$ (g cm$^{-2}$) & $p$ & rot & $t_{\rm end}$ \\ \hline
1 & 50 AU  & $0^{\circ}$  & 170 & 1.5 & pro    & 150 kyr \\
2 & 50 AU  & $0^{\circ}$  & 170 & 1.5 & retro  & 150 kyr \\
3 & 50 AU  & $35^{\circ}$ & 170 & 1.5 & retro  & 150 kyr \\
4 & 50 AU  & $60^{\circ}$ & 170 & 1.5 & retro  & 150 kyr \\
5 & 50 AU  & $90^{\circ}$ & 170 & 1.5 & ---    & 150 kyr \\
6 & 50 AU  & $0^{\circ}$  &  17 & 1.5 & pro    & 150 kyr \\
7 & 50 AU  & $0^{\circ}$  &  17 & 1.5 & retro  & 150 kyr \\
8 & 50 AU  & $90^{\circ}$ &  17 & 1.5 & ---    & 150 kyr \\
9 & 50 AU  & $90^{\circ}$ & 170 & 1   & ---    & 100 kyr \\
10 & 50 AU  & $35^{\circ}$ & 170 & 1.5 & pro  & 150 kyr \\
11 & 50 AU  & $60^{\circ}$ & 170 & 1.5 & pro  & 150 kyr \\

\end{tabular}
\caption{Summary of the setups of the different runs. First column: label of the run, column two: initial inner disk size, column three: inclination angle of infalling cloudlet $\alpha$, fourth column: column density $\Sigma_0$ of inner disk, fifth column: exponent of inner disk profile $p$, sixth column: rotation of inner disk with respect to infalling cloudlet (pro for prograde, retro for retrograde), seventh column: end of simulation. }
\label{runover}
\end{table}

\begin{figure*}
    \includegraphics[width=\textwidth]{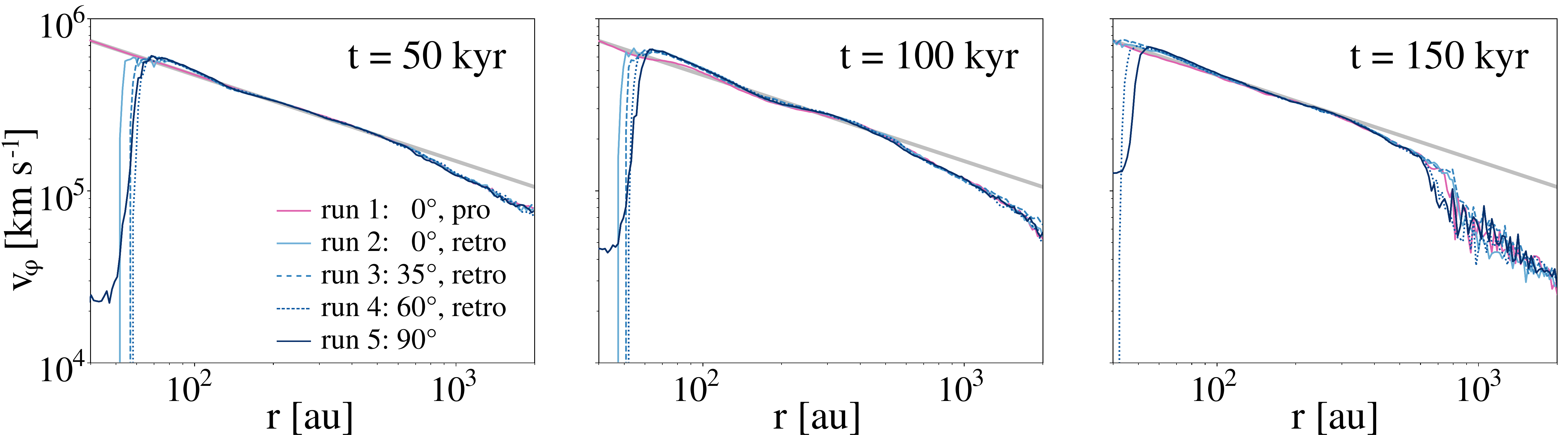} 
            \caption{Average azimuthal velocity $v_{\phi}$ profile of the outer disk for run 1 (purple solid line) and run 2 (solid), run 3 (dashed), run 4 (dotted), run 5 (solid) at $t=50 \unit{kyr}$ (left panel), $t=100 \unit{kyr}$ (middle panel) and $t=150 \unit{kyr}$ right panel). 
            The darker the color of a line, the larger the inclination angle $\alpha$. 
            $v_{\phi}$ is computed within a vertical extent of $|z| < \mathrm{max}(6 \unit{AU}, 0.05 r)$ from the midplane of the disk for 300 logarithmically spaced radial bins in the range from $10$ to $10^4 \unit{AU}$. 
            The grey line is the Keplerian profile for a $2.5 M_{\odot}$ star.}
        \label{fig:r_vphi_out}
\end{figure*}

\begin{figure*}
    \includegraphics[width=\textwidth]{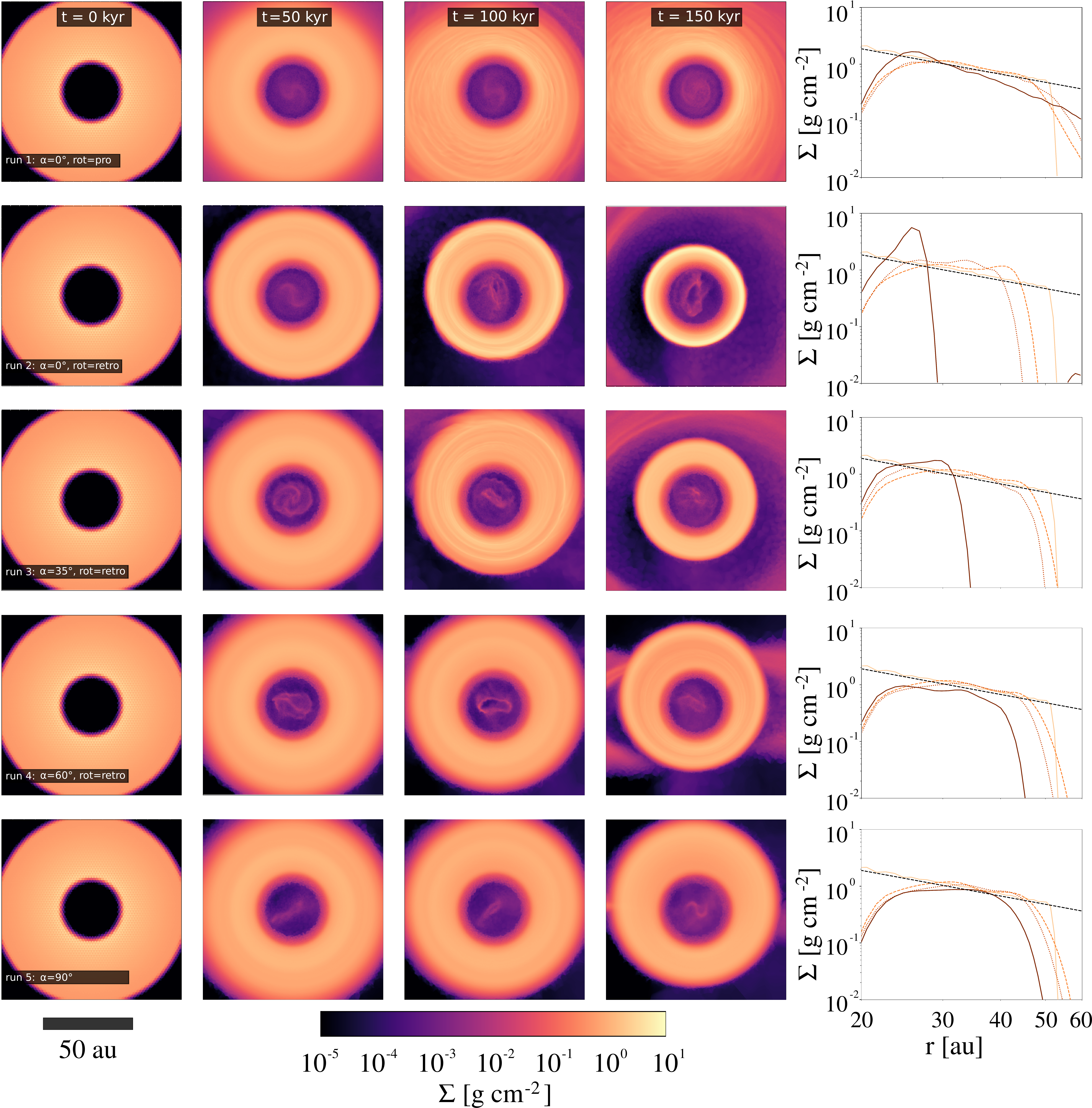} 
        \caption{Projection of the inner disk for run 1 to 5 (top to bottom) at $t=0$, $50$, $100$ and 150 kyr (from left to right) measured perpendicular to the initial plane of the inner disk and azimuthally averaged column density profile at the shown snapshots (rightmost panels). 
        The azimuthally averaged column density in the rightmost panels is computed within a vertical extent of $z=\pm 6 \unit{AU}$ and the darkness of the line corresponds to the time of evolution (from light orange representing $t=0 \unit{kyr}$ to dark orange at the end of the simulation). Note that the inner part is affected by accretion onto the central star and gravitational softening. We therefore exclude the inner region within $r=r_{\rm acc}+5 \unit{AU}=r_{\rm grav,soft}+10 \unit{au}=20 \unit{AU}$ in the azimuthally averaged column density profiles.}
        \label{fig:1p5_100_seq}
\end{figure*}

\begin{figure*}
    \includegraphics[width=\textwidth]{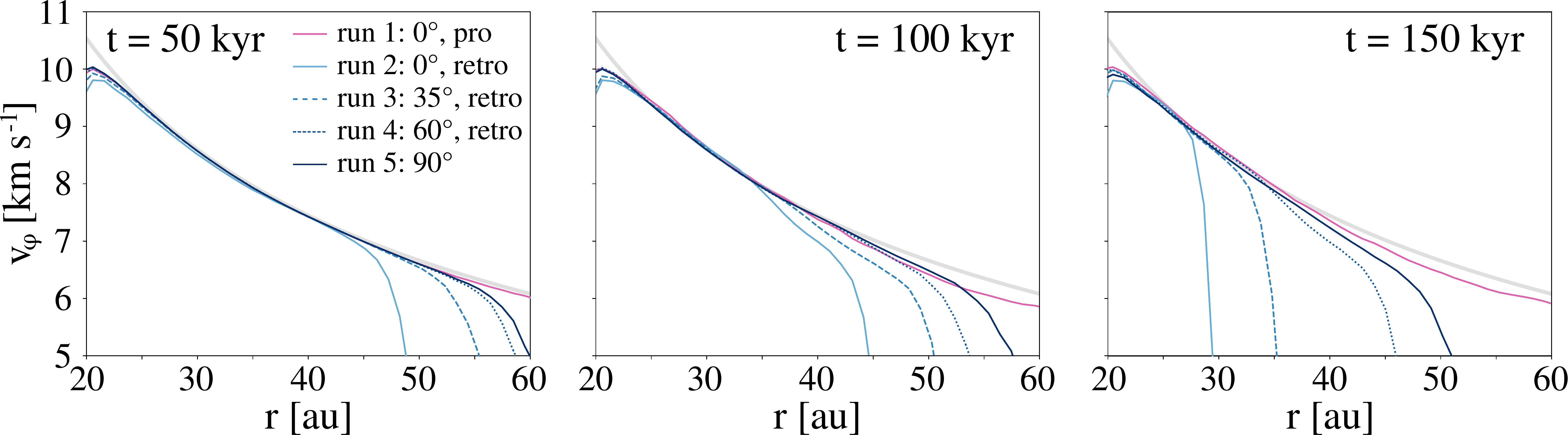}
            \caption{Similar as \Fig{r_vphi_out}, but for the inner disk. $v_{\phi}$ is computed within a vertical extent of $|z| < 6 \unit{AU}$ from the midplane of the disk for linearly spaced radial bins of size $\Delta r = 2 \unit{AU}$. }
        \label{fig:r_vphi_in}
\end{figure*}

\section{Results}
For illustration purposes we show the evolutionary sequence on spatial scales of $\pm4000$ AU, $\pm800$ AU and $\pm160$ AU around the central star for run 2 and 5 in \Fig{1p5_seq_000} and \Fig{1p5_seq_090}. (For comparison, the sequences for run 1, 3, 4, 10 and 11 are shown in appendix A.) 

\subsection{Outer disk}
The figures show that in all cases the infalling cloudlet leads to the formation of a second-generation disk associated with larger-scale spiral structures for all of our runs. 
The outer disk extends up to $\sim 1000$ AU. 
During the first several 10 kyr of each simulation, the inner disk is already in a stable state that is close to equilibrium, while the outer disk is in its formation phase.
In this initial phase of outer disk formation, the second-generation disk evolves on smaller time scales than the inner disk although the orbital and hence dynamical time $t_{\rm orb}$ of the inner disk is significantly shorter due to the scaling of $t_{\rm orb} \propto R^{3/2}$. 
Clear signatures of the dynamical nature are the arc structures that form during the encounter events and extend to distances beyond $1000$ AU in good agreement with the results found in KGD20. In all cases, the outer disk forms with an inclination to the primordial disk equivalent to the initial infall angle $\alpha$ of the individual runs.

To quantify the velocity profile of the outer disk, we compute the azimuthal velocity $v_{\phi}$. 
The azimuthal velocity is calculated perpendicular to the angular momentum vector of the infalling cloudlet with respect to the central star given by $\mathbf{L} = M_{\rm cloudlet} (\mathbf{r_{\rm init, cloudlet}} \times \mathbf{v_{\rm cloudlet}})$.
As expected from the results in KGD20, the velocity structure of the outer second-generation disk is close to Keplerian for cases 1 to 5 (see \Fig{r_vphi_out}) to radial distances of several 100 AU. 
The profiles only marginally differ for the different infall angles $\alpha$. 
Comparing the radial extent of the Keplerian profile at different points in time shows that the outer disks become more compact over time and the density drop at the outer edge of the second-generation disk becomes more distinct. However, the azimuthal velocity still is approximately Keplerian up to radial distances of at least 300 AU for all cases at $t=150 \unit{kyr}$. 
At the inner edge, the azimuthal velocity profile extends further inwards at the end of the simulation ($t=150 \unit{kyr}$) for the runs with smaller infall angles $\alpha$.    
The Keplerian velocity profile continues to small radii in the case of $\alpha=0$ and initial parallel orientation of the cloudlet's angular momentum vector and inner disk (run 1).

\subsection{Inner disk}
The properties of the inner disk are more affected by the combination of infall angle $\alpha$ and the orientation of the rotation axis of the inner disk. 
\Fig{1p5_100_seq} shows face-on projections for run 1 to 5 at $t=0 \unit{kyr}$, $t=50 \unit{kyr}$, $t=100 \unit{kyr}$ and $t=150 \unit{kyr}$, as well as the azimuthally averaged radial column density profile of the inner disk. 
After the start of the simulation, mass located within the accretion radius $r=15 \unit{AU}$ accretes onto the central star. As not all of the mass within the accretion radius is immediately accreted, temporary features such as rings and small spirals of low-density form within the accretion radius. 
However, as the dynamics of the gas within $r_{\rm acc}=15 \unit{AU}$ are strongly affected by the numerical parameters of the accretion recipe and gravitational softening, we exclude this region from the analysis.
Accretion also leads to a drop in column density $\Sigma$ up to a radial distance of $\approx 20$ AU to $\approx 25$ AU from the central star. 
However, in all cases, the slope of the inner disk remains similar to the initial setup between $\approx 25$ AU and $\approx 50$ AU during the first $t\approx 50 \unit{kyr}$. For the disks with lower density (runs 6 to 8), the inner disk starts to align with the outer disk, while the inner disk remains in its original orientation for the most massive inner disk (run 9). We investigate this aspect in more detail in section 3.5.      

The size of the inner disk remains similar for the case of prograde infall with $\alpha=0$, while the disk becomes smaller for the case of retrograde infall with $\alpha=0$. 
The inner disk in the runs with retrograde infall at low $\alpha$ becomes more compressed with a bump in the density profile at $r\approx 25 \unit{AU}$.
Due to angular momentum exchange between outer and inner disk, the inner disk loses angular momentum, and thus the gas moves inward and piles up. 
Consistently with the transport of angular momentum, more mass accretes onto the central star in the retrograde scenario compared to the prograde scenario (see section 3.3). 
In contrast, the density profile changes less compared to the initial profile for increasing inclination angle $\alpha$, as well as for prograde infall.
The azimuthal velocity profile of the inner disk shows a similar pattern (see \Fig{r_vphi_in}).
The velocity profile is Keplerian beyond the initial size of the inner disk for the case of prograde, non-inclined infall.
For retrograde infall, the radial extent of the Keplerian profile decreases over time for all cases, but the radial extent shrinks more quickly for lower $\alpha$ than for larger $\alpha$. 

Both the density and velocity profiles of the retrograde infall runs 2 to 5 show a clear trend: the higher the infall angle $\alpha$, the less affected is the size of the inner disk.
This trend is expected as the angular momentum of the infalling cloudlet is maximal opposite to the angular momentum of the inner disk in run 2 (retrograde infall without inclination $\alpha$), and hence brakes the inner disk. 
An increasing inclination angle $\alpha$ weakens this effect to the point at $\alpha=90^{\circ}$ where the angular momentum vectors do not oppose each other in the disk plane.
Consistently, the extent of the inner disk is largest at $t=150 \unit{kyr}$ in the case of prograde infall, where outer and inner disk form in the same plane along the same rotational axis.    

During the shrinking phase of the inner disk, the transfer of angular momentum also triggers the formation of strong ring features in the inner disk. 
The ring features are less deep for prograde than for retrograde infall.
Again, the analysis shows a correlation with inclination. 
The larger the inclination angle with respect to the primordial disk plane, the lower is the amplitude of the forming rings. 
For run 5 ($\alpha=90^{\circ}$), even after $t=150$ kyr the density distribution of the inner disk is still very smooth, while the disk in run 2 shows the strongest ring structures preceding the formation of the smallest disk towards the end of the simulation. 

\subsection{Counter- vs co-rotating disks}
To provide a better understanding of an infalling event on the inner disk, we analyze the two extreme cases of perfectly prograde and perfectly retrograde infall of run 1 and 2 in more detail. 
Generally, run 2 shows the important result that two counter-rotating disks can form as a consequence of late infall events if the infalling material accretes with an opposite angular momentum orientation. 
Consistently, the infalling cloudlet in run 1 leads to the formation of a second-generation disk with the same rotation axis as the primordial disk. 
These results are consistent with results from hydrodynamical simulations by other groups, who find the formation of counter-rotating disks in case of retrograde infall during the early collapse phase \citep{Vorobyov2015, Bate2018}.  
Apart from the opposite orientation, the spiral-like features on larger scales as well as the radial extent of the second-generation disk are similar in both scenarios. 

\begin{figure}
    \includegraphics[width=\columnwidth]{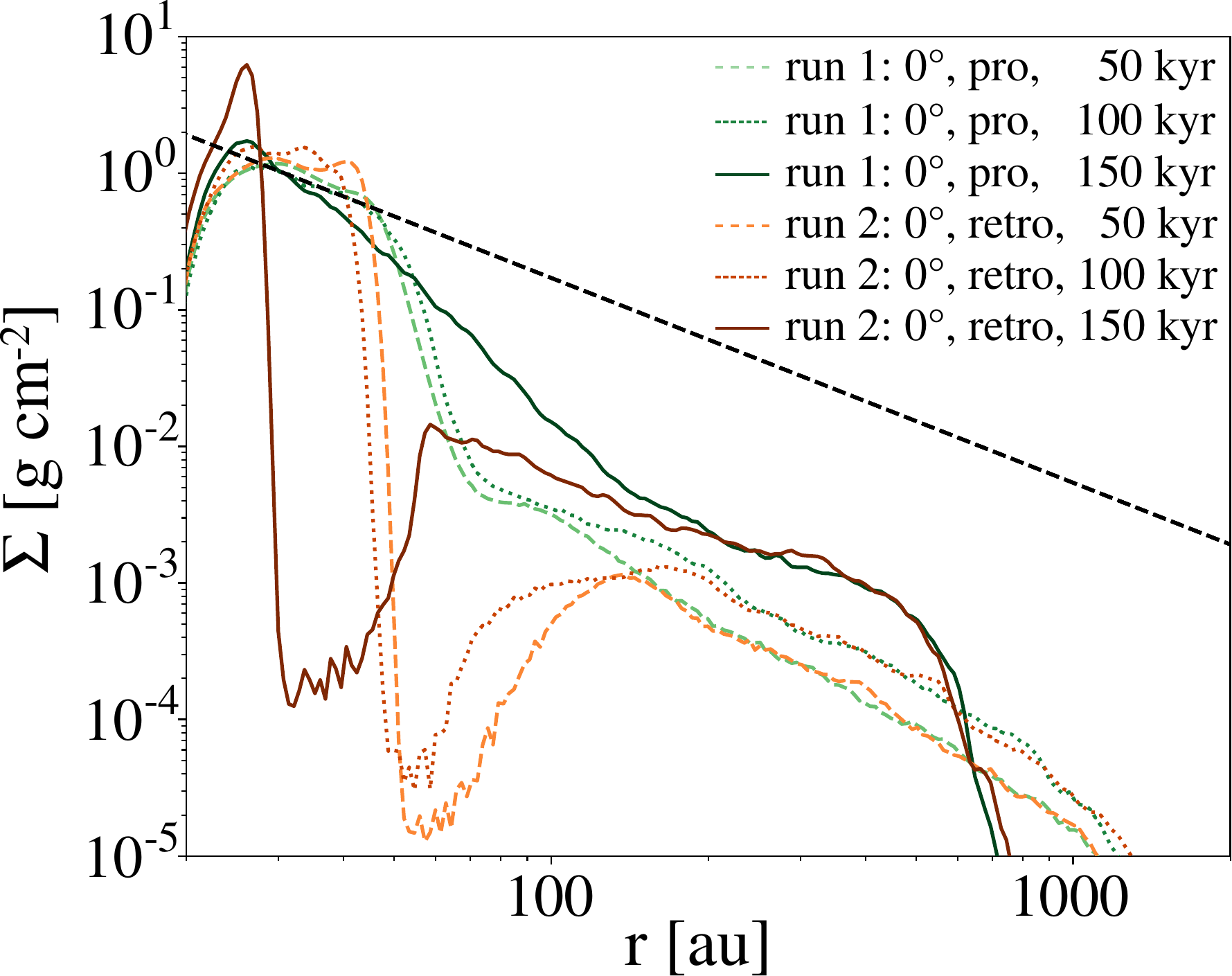} 
        \caption{Column density $\Sigma$ over radius $r$ for prograde infall without inclination angle in run 1 (green colors) and retrograde infall without inclination angle in run 2 (orange colors) at $t= 50 \unit{kyr}$ (dashed lines), $t=100$ kyr (dotted lines) and $t=150$ kyr (solid lines). 
        The darkness of the line colors indicates the evolution in time. From 50 kyr (lightest color) to the final time (darkest color) at $t=150 \unit{kyr}$. 
        The black dashed line corresponds to a theoretical column density profile for $\Sigma_0=170 \unit{g}\unit{cm}^{-2}$ and $p=1.5$ that extends to 2000 AU.  }
        \label{fig:r_Sigma_alpha_000}
\end{figure}

On smaller scales, i.e. on scales of the transition between outer and inner disk, we can see substantial differences between the two extreme scenarios.
The transition between outer and inner disk is smooth in the aligned infall case. 
In contrast, outer and inner disk in run 2 are more distinctly separated by a transition zone of $\sim 10 \unit{AU}$ in width. 
As shown in \Fig{r_Sigma_alpha_000}, the gap is about 30 AU in radial width, while the inner disk has shrunk to about 25 AU in radius for run 2 at $t=150 \unit{kyr}$. 
The inner disk in run 2 shrinks more than in run 1 during the evolution. 
In the misaligned cases (run 3, 4 and 5), the gap is smaller than in the retrograde scenario without inclination, but still more than $\approx 10 \unit{AU}$, while the size of the inner disk shrinks less as explained above.
The features are a consequence of the opposite orientation of the angular momentum vector of the outer disk \citep[see also][]{Wijnen2016,Wijnen2017b}. 
The transition zone in form of a gap forms where the gas of the forming outer disk interacts with gas of the inner disk \citep[see also][]{Vorobyov2016}. 
During the interaction, the outer part of the inner disk transfers part of its angular momentum to the inner part of the outer disk. 
As a consequence, the inner disk shrinks in size, and hence a gap forms in between inner and outer disk.

Gaps are an important feature of transition disks.
In the majority of cases, the gap size is larger than the size of the inner disks. 
In our models, the deepest and largest gap occurs for retrograde infall, where the gap size is about 30 AU and exceeds the radius of the shrinking inner disk. 
By comparing simulations of counter-rotating disks with simulations of disks with an embedded giant planet, \citet{Vorobyov2016} showed that the resulting gaps are similar. 
Therefore, infall-induced gaps in counter-rotating disks might be mistakenly interpreted as signs of a planet-bearing disk.
Searching for tracers of accretion onto a giant planet can help to reveal the gap origin, such as done with H$\alpha$ emission for PDS 70 \citep{Keppler2018}. 
In the majority of cases in our simulations, the gap region is smaller than the inner disk. 
Note that the initial size of 50 AU for the inner disk is a somewhat arbitrary choice in our setup and we expect larger ratios of gap size to inner disk radius for smaller initial inner disk sizes and/or larger impact parameters $b$.

Generally, the possibility of angular momentum transport from inner to outer disk means that material moves radially inwards, and hence we expect an increased accretion rate for the counter-rotating case compared to the aligned case as shown theoretically \citep{Lovelace_Chu1996} and in earlier hydrodynamical simulations \citep{Kuznetsov1999} for counter-rotating disks.
To test whether this is the case in our simulation, we measure the accretion rate as the amount of mass that accretes onto the central sink particle per unit time.

\begin{figure}
    \includegraphics[width=\columnwidth]{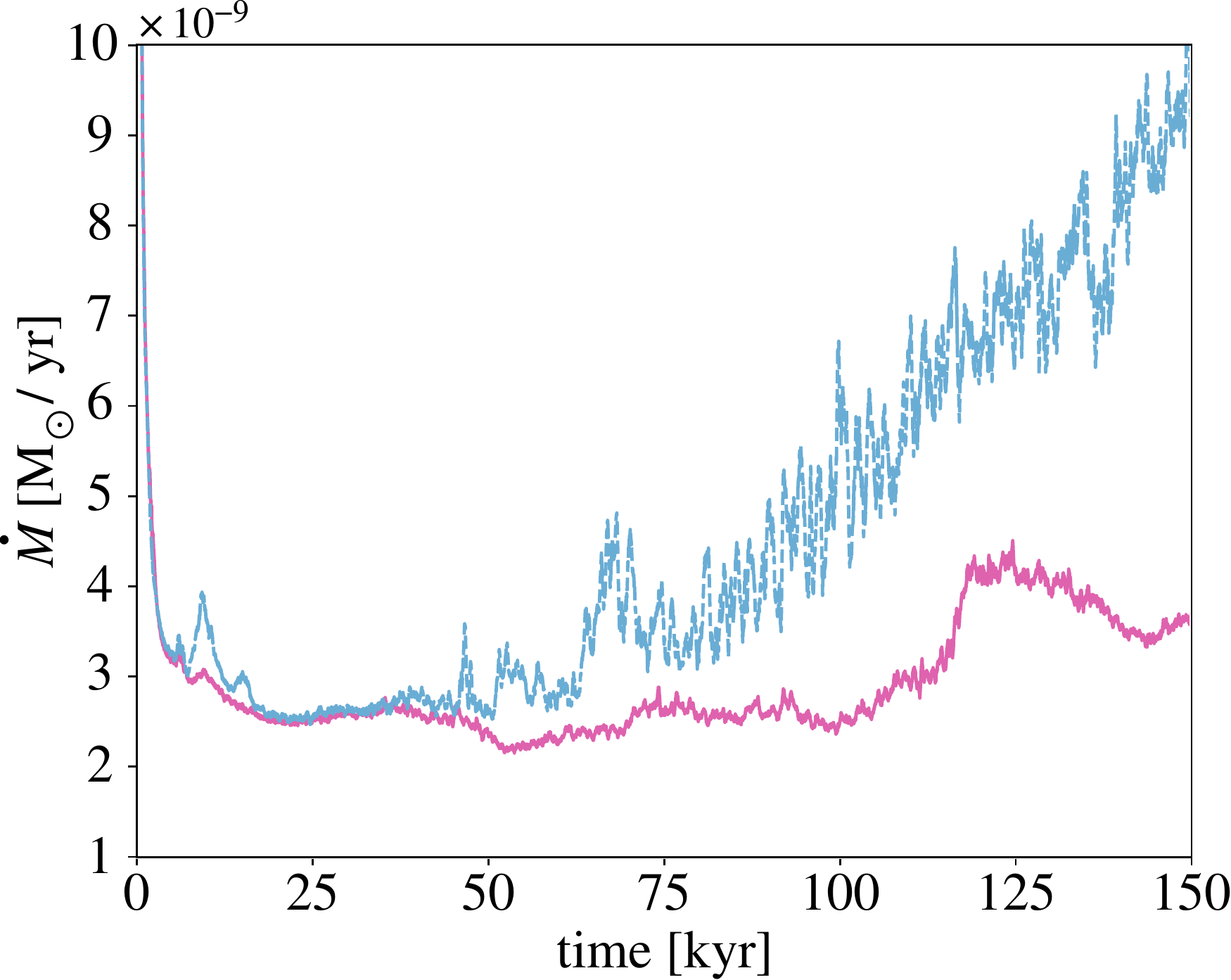} 
        \caption{Time evolution of accretion rate for aligned infall run 1 (purple line) and counter rotating infall run 2 (blue line) with infall angle $\alpha=0^{\circ}$.}
        \label{fig:macc}
\end{figure}

\Figure{macc} shows $\dot{M}$ of run 1 and 2, that are the two extreme cases of prograde and retrograde infall without inclination. 
In agreement with expectation, the accretion rate increases to higher values in the retrograde infall run after $t\approx 50 \unit{kyr}$, while $\dot{M}$ remains smaller in the run with prograde infall.
The delay between the visible increase in the accretion rate compared to the first contact of the encountering material with the inner disk can be understood as the time it takes to propagate the angular momentum transfer from the outer part of the inner disk to the vicinity of the star.

The column densities in the outer disk are significantly lower than in the inner disk as expected from the significant mass difference of inner disk and infalling cloudlet mass. 
For runs 1 to 5, the initial total mass of the infalling cloudlet is less than $25 \%$ of the initial inner disk mass.    
This can be seen particularly well in \Fig{r_Sigma_alpha_000}, which shows the radial profile of column density for run 1 and 2, i.e., the scenario in which the outer disk forms in the same plane as the inner disk.
Similar to the inner disk, the column density profile of the outer disk follows a $\Sigma \propto r^{-1.5}$ relation extending from less than 200 AU to about 1000 AU. 
During the 150 kyr of evolution the slope remains similar to $\Sigma \propto r^{-1.5}$, 
while the overall profile is shifted to increasing column densities. 
As shown in the velocity profile (\Fig{r_vphi_out}), the outer disk becomes more compact and distinct at its outer edge toward the end of the simulation.

\begin{figure}
    \includegraphics[width=\columnwidth]{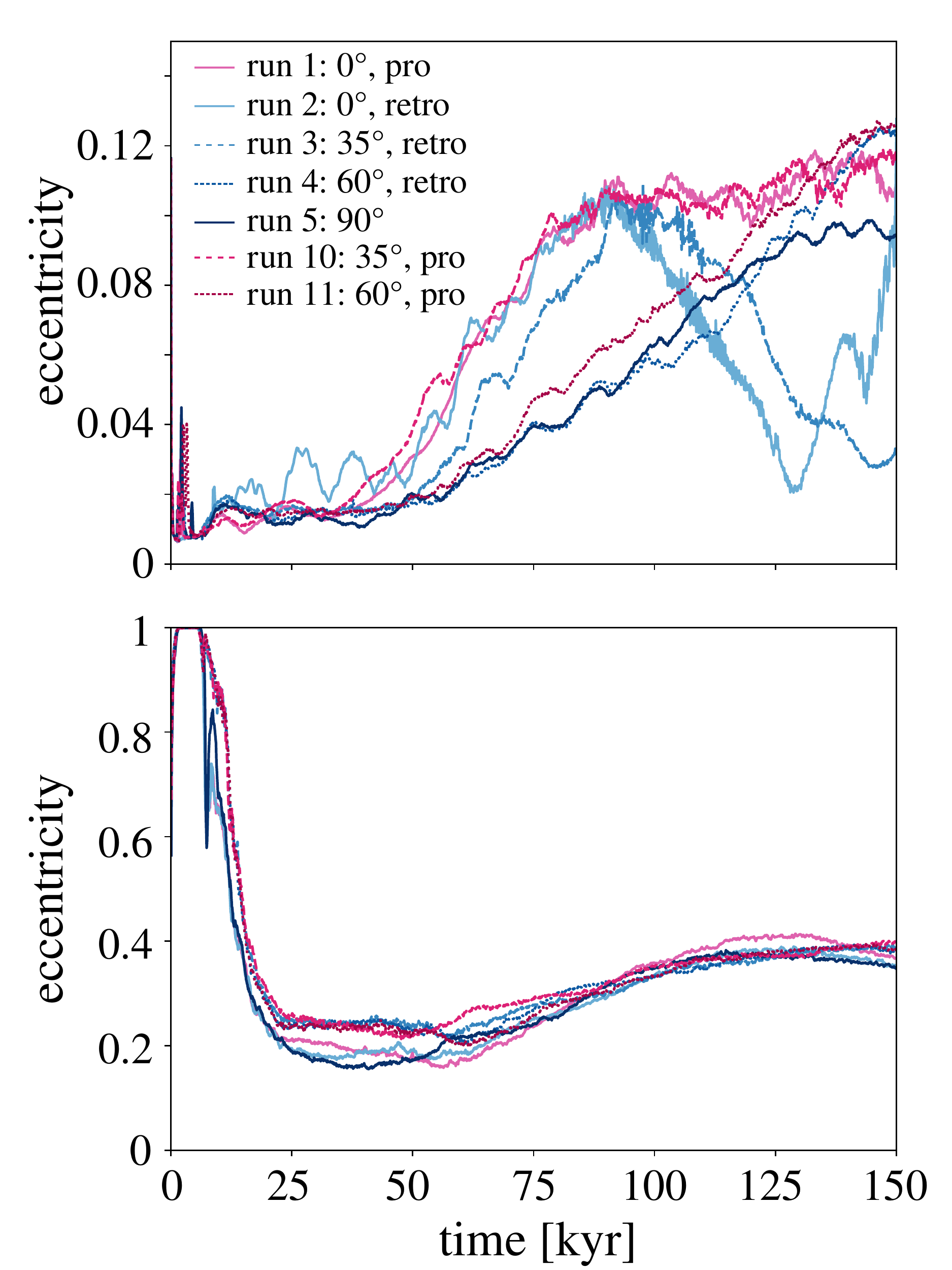} 
        \caption{Time evolution of azimuthally and radially averaged eccentricity in inner (top panel) and outer disk (bottom panel). For the inner (outer) disk, the average eccentricity between a radial distance of $r=25$ (200) AU to 30 (225) AU and a vertical extent of $z=\pm 0.1 r$ over the entire azimuth is shown. }
        \label{fig:ecc}
\end{figure}

\begin{figure*}
    \includegraphics[width=\textwidth]{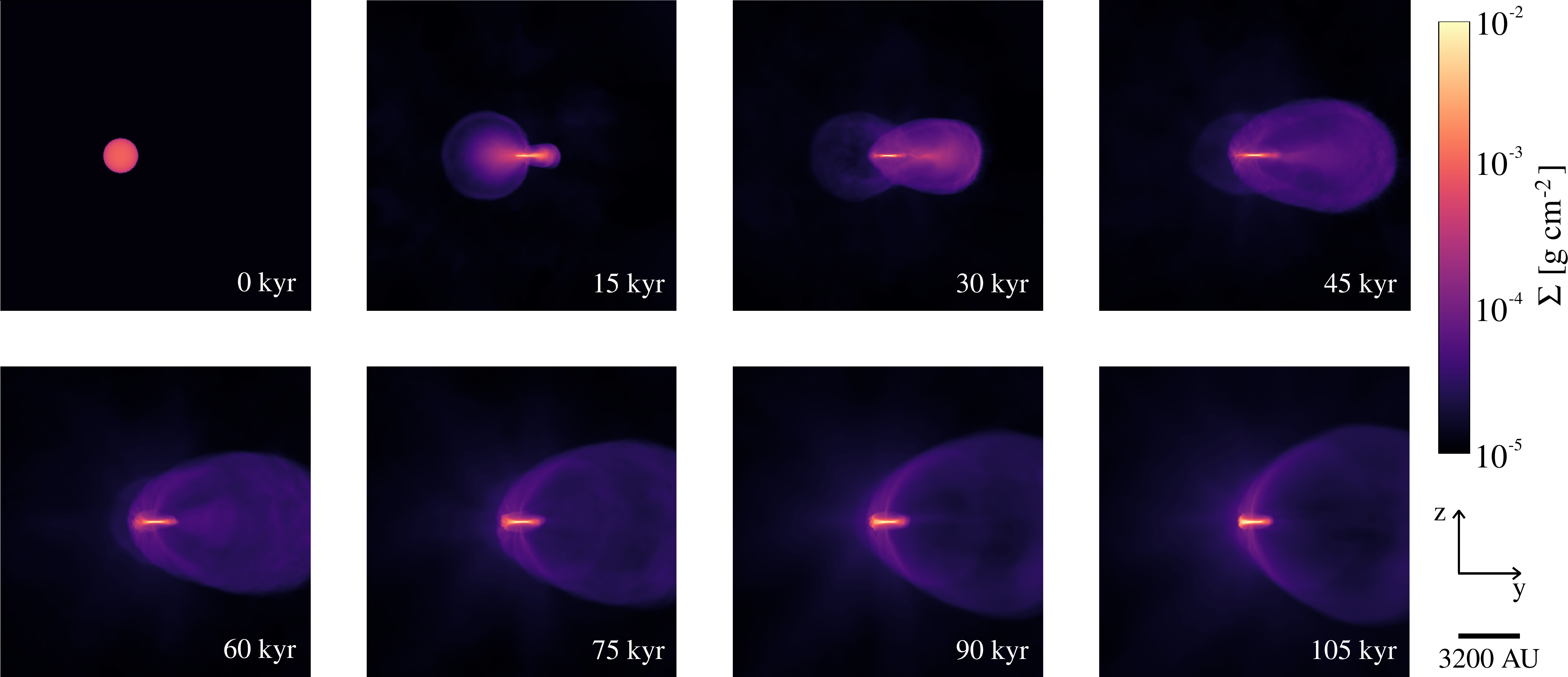} 
        \caption{Evolution of run 2 seen along the x-axis for 105 kyr showing the column density every 15 kyr. The images show how the cloudlet expands and how gas falls towards the central star forming the disk while the cloudlet dissolves.}
        \label{fig:seq_infall_run2}
\end{figure*}

\begin{figure*}
    \includegraphics[width=\textwidth]{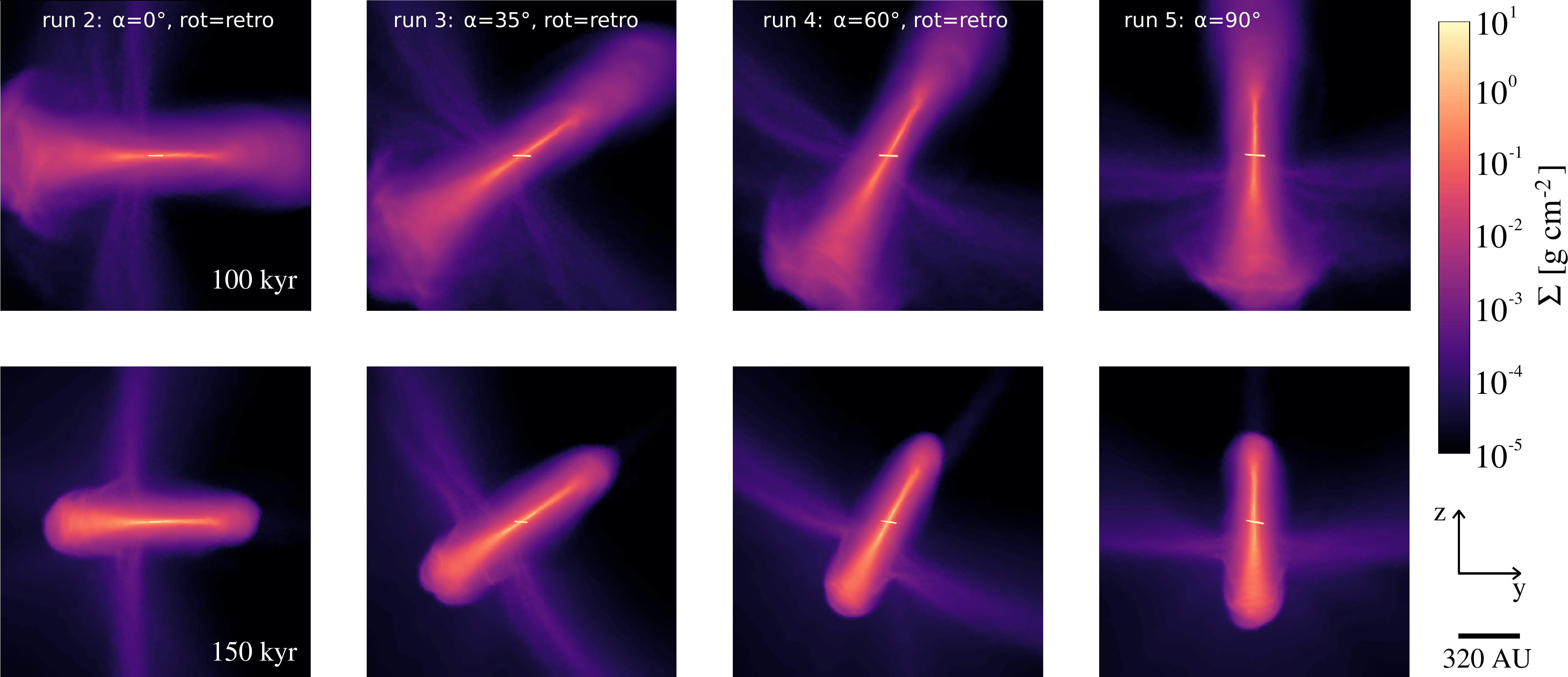} 
        \caption{Inner and outer disk of run 2 to run 5 (from left to right panel) seen from a viewing angle along the x-axis at $t=100 \unit{kyr}$ (upper panel) and $t=150 \unit{kyr}$ (lower panel).  }
        \label{fig:evo_yz}
\end{figure*}

\begin{figure}
    \includegraphics[width=\columnwidth]{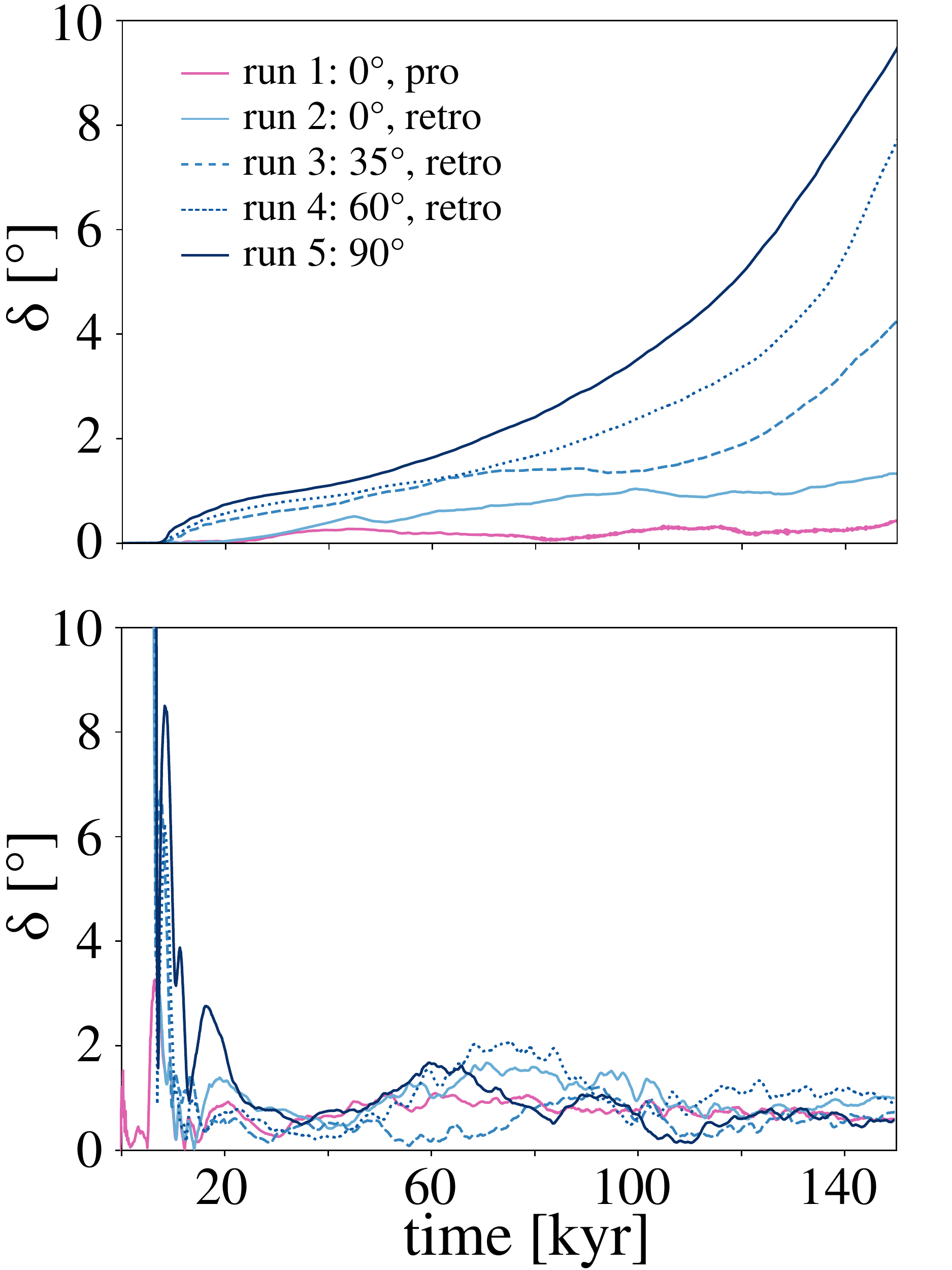} 
        \caption{Time evolution of angle $\delta$ for inner (top panel) and outer disk (bottom panel). For the inner disk, $\delta$ is measured as the angle of the angular momentum vector at time $t$ with respect to the initial angular momentum vector of the inner disk $\mathbf{L}_{50}(0)$. For the outer disk, $\delta$ is the angle between the angular momentum vector of gas located within a range from radius $r=200$ AU to $r=225$ AU and the initial angular momentum vector of the infalling cloudlet.  }
        \label{fig:delta}
\end{figure}

\subsection{Eccentricity}
The infall onto the primordial disk is asymmetric. 
As a consequence both outer and inner disk become eccentric as seen in the projection plots shown in \Fig{1p5_seq_000}, \Fig{1p5_seq_090} and \Fig{1p5_100_seq}.
To quantify this aspect further, the eccentricity $e$ is computed for each grid point according to
\begin{equation}
    e = \sqrt{1 - \frac{2 |\epsilon| h^2}{(GM_{\rm tot})^2}},
\end{equation}
where $\epsilon$ is the specific orbital energy
\begin{equation}
    \epsilon = \frac{E_{\rm pot} + E_{\rm kin}}{\mu_{\rm red}}
\end{equation}
$h$ is the specific angular momentum 
\begin{equation}
h = \frac{L}{\mu_{\rm red}},    
\end{equation}
where $\mu_{\rm red} = \frac{M_{*} m}{M_{\rm tot}}$ is the reduced mass with $m$ being the mass of the individual grid cell and $M_{\rm tot}=M_{*} + m \approx M_{*}=2.5 M_{\odot}$.
In \Fig{ecc} we show the time evolution of eccentricities in the inner and the outer disk. 

In all cases, the forming second-generation outer disk is significantly more eccentric than the initially circular primordial disk. 
The time evolution of the eccentricities of the outer disk illustrates the sequence of the encounter event. 
At the beginning of our run, the outer disk has not formed yet, which is why the eccentricity is close to its maximum value at 1. 
When the cloudlet approaches the central star, part of the gas is accreted by the central star causing a rapid decrease in eccentricity to $e\approx 0.2$ from $t\approx 10$ kyr to $t\approx 20$ kyr. 
The eccentricity remains at $e\approx 0.2$ for about 50 kyr before it rises again up to $e\approx 0.4$ at $t=110$ kyr. 
The rise in eccentricity occurs when gas that initially passes by the star with high speed returns to the scales of the outer disk for a second time.   
This pattern is similar for all runs though there are some minor differences between the setups. 
The runs with intermediate angles, i.e., 35$^{\circ}$ and 60$^{\circ}$ remain slightly more eccentric between 20 to 70 kyr than the other three runs with either perpendicular infalling angle or zero inclination angle. 
After $t=110$ kyr, the disk in the runs with intermediate inclination remain eccentric with $e \approx 0.4$, while the eccentricity slowly decreases again for the other runs. 
These differences are minor and not substantial though.

The inner disk remains in all cases less eccentric than the newly forming second-generation disk.  
Comparing the eccentricity in the inner disk of the different runs, however, shows differences.
The eccentricity rises more rapidly from an initially circular orbit for the runs without any inclination angle. 
Also, the eccentricity rises more quickly for the run with low inclination of 35$^{\circ}$ compared to the runs with higher inclination.
Additionally, for the runs with zero or low inclination angles, the maximum eccentricity is highest with $e> 0.1$. 
However, after the eccentricity is at its maximum after $t\approx 90$ kyr, the inner disk becomes less eccentric again for the low or zero inclination {retrograde cases, while it remains above $e\approx 0.1$ for the prograde cases. }
Note that in the case of retrograde infall at zero inclination angle (run 2), the drastic increase in eccentricity shown at radial distance between 25 to 30 AU after $t=130$ kyr is not associated with the disk anymore, but mostly traces the low-density gas beyond the shrinking inner disk (see evolution of the disk in \Fig{1p5_100_seq}).  
The reduction in eccentricity in the retrograde low inclination runs correlates with the compression of the inner disk. In contrast, $e$ continuously increases for the high inclination cases for both retro- and prograde infall until the end of the simulation to $e\approx 0.12$.
Comparing the retrograde and prograde scenarios at identical infall angles, we find that the rise in eccentricity occurs on average slightly more rapidly for prograde infall than retrograde infall. 
This is likely also related to the fact that for retrograde infall more mass is accreted onto the central star, while in the prograde scenarios more mass remains in the domain and contributes to the higher eccentricity.

\subsection{Gas infall and orientation of inner with respect to outer disk over time}
As the cloudlet is not pressure confined it starts to expand from $t=0$ \cite[see \Fig{seq_infall_run2} and also discussion in][]{Dullemond2019}. 
During the cloudlet encounter with the central star, part of the gas is captured by the central gravitational potential and forms the outer disk. The remaining part is only deflected. 
The part of the expanding remnant of the cloudlet  furthermost from the star continues to expand, escapes the gravitational pull from the central star moves away from the star-disk system. 
The part of the remaining cloudlet that is closer to the central gravitational potential, but not yet located in the second-generation disk eventually falls toward the star-disk system. 
Part of the infalling gas at the border of the highly disrupted cloudlet appears as a faint filament oriented almost perpendicular with respect to the disk plane when looking at the system on scales of $\sim 1000 \unit{AU}$ along the x-axis (see \Fig{evo_yz}).

As a consequence of the non-zero inclination angle with respect to the disk plane in the runs 3, 4 and 5, the outer disk forms with the corresponding inclination to the inner disk as illustrated in \Fig{evo_yz}. 
The outer disk does not show any warp during the time evolution of the eleven runs. 
As the gas in the primordial inner disk and the newly forming outer disk exchange angular momentum, the orientation of the inner disk can change during the simulation. 
Therefore, we compare the orientation of the angular momentum vector of the inner part at different times with each other. In particular, we compute the angular momentum vector $\mathbf{L}_{50}(t)$ for all the gas located within a radial distance $r$ of $15 \unit{AU} < r < 50 \unit{AU}$ from the central star.
\Fig{delta} shows the evolution of the angle between the angular momentum vector at each snapshot with respect to the initial orientation of the angular momentum vector $\mathbf{L}_{50}(0)$. 
The plot shows that the inner part of the gas undergoes mild changes in all cases. 
Consistently, infall with higher inclination angles generally correlates with higher deviation angles from the initial state of the inner disk.  

\begin{figure}
    \includegraphics[width=\columnwidth]{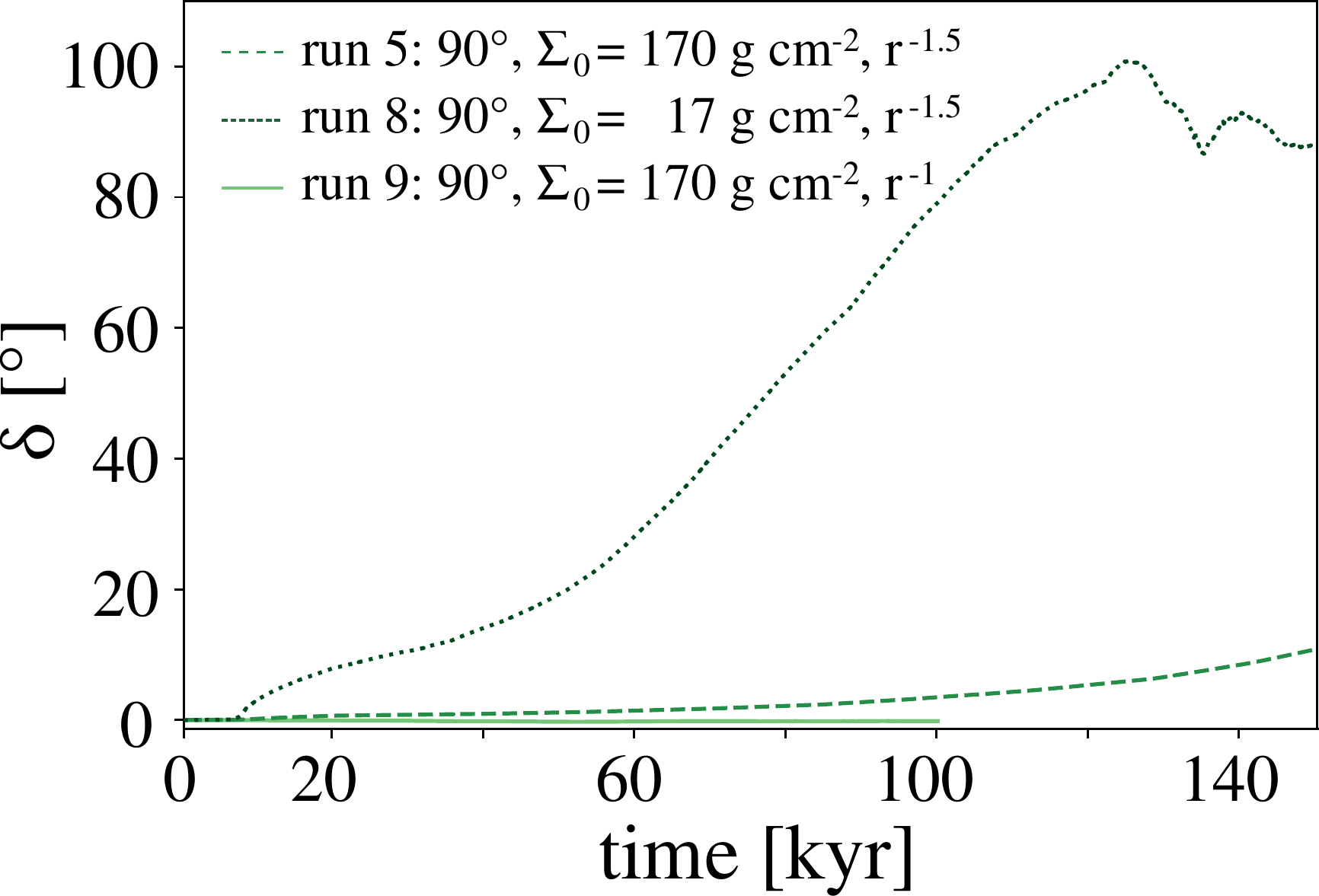} 
        \caption{Same as upper panel in \Fig{delta}, but for runs 5 (dashed line), 8 (dotted line) and 9 (solid line) only, i.e., the runs with infall angle $\alpha=90^{\circ}$, and varying initial density profile of the inner disk. 
        The initial inner disk mass is $M_{\rm i,disk}\approx 4 M_{\rm cloudlet}$ in run 5, $M_{\rm i,disk}\approx 0.4 M_{\rm cloudlet}$ in run 8 and $M_{\rm i,disk}\approx 24 M_{\rm cloudlet}$ in run 9. In all cases $M_{\rm cloudlet} = 1.87 \times 10^{-4} \unit{M}_{\odot}.$}
        \label{fig:delta_090}
\end{figure}

For the main runs, the disks remain close to their initial state by the end of the simulation. Even in the case of 90 degree infall $\delta$ is less than 12 degrees by the end of the simulation at $t=150$ kyr. 
However, the inner disk becomes continuously more tilted for the runs with higher inclination angles suggesting that they will eventually align with the outer disk. 
The inner disk aligns with the second-generation outer disk, as the angular momentum of the outer disk exceeds the angular momentum of the inner disk.
The comparison run of infall angle $\alpha=90^{\circ}$ with ten times less mass in the inner disk (run 8), hence ten times less angular momentum, as well as the run with a more massive inner disk (run 9) confirm this result (see \Fig{delta_090}). 
For the low-density disks, the inner disk starts to align with the outer disk earlier. 
By the end of the simulation at $t=150 \unit{kyr}$, the inner disk is almost fully aligned with the outer disk. 
In contrast, the more massive disk in run 9 is still oriented close to its initial orientation at the end of the simulation at $t=100 \unit{kyr}$. 

This also suggests that in configurations where the inner disk is massive enough that it exceeds the angular momentum of the forming outer disk, the outer disk would align with the orientation of the inner disk. 
We did not run models with such configurations, but it remains an important study for the future. 
Especially, the question whether the outer disk becomes (temporarily) warped in a setup of continuous infall is intriguing.
Comparing the results of such a scenario with existing work on warped disks induced by a perturbing companion \citep{Papaloizou1995,Larwood1996,LubowOgilvie2000,Martin2019} will be particularly relevant given the observations of warped disks around young stars.
In our models, the inner disk changes orientation as a whole, but does not become warped. 
This is consistent with previous models of small disks with low viscosity (wave-like regime), where the disk globally bends as a consequence of a warp wave that is relatively large compared to the disk \citep{NixonPringle2010,NixonKing2016}. 
In cases of higher viscosity (diffusive regime), warping and even tearing of the disk would happen more easily (see section 4.3).

\section{Discussion and future prospects}

\subsection{Limitations of the model}
The key question we address in this paper is whether misaligned disks can form as a result of a late infall event. 
To test this possibility, the model setup is idealized.
The central star only acts as a point mass and the infalling material is initially confined in a spherical cloudlet. 
In our model the gas is assumed to be isothermal and cold with a temperature of 10 K.
In reality, the inner disk has a temperature profile due to heating by the central star. 
A more realistic treatment of the thermodynamics would change the properties of the inner disk, e.g., causing flaring of the disk,
but we showed in KGD20 that heating from the central star only marginally affects the formation of the second-generation disk due to its distance to the heating source.

Another simplification of our model is the lack of magnetic fields.
Magnetic fields are supposed to play a major role in transferring angular momentum during the protostellar collapse phase, in particular through the launching of outflows from the inner disk.
At later times, magnetic fields are likely to be less important for more evolved disks as the field strength dissipates over time. 
Magnetic fields likely affect the younger second-generation disk by reducing its disk size via magnetic braking. 
The corresponding transfer of angular momentum would hence lead to stronger shrinking of the inner edge of the outer disk and thereby indirectly affect the size of the inner disk.
Nevertheless, given the high angular momentum of an encounter event for sufficient impact parameter $b$, we do not expect quenching of infall-induced rotational structures when incorporating magnetic fields.

There is one particular aspect regarding the role of magnetic fields that would be interesting to test in future works though. If magnetic braking transfers enough angular momentum from the outer disk, the angular momentum budget of the outer disk can become less than the angular momentum of the inner disk. 
As a consequence, the inner disk would cause a stronger torque on the outer disk, and the outer disk would align with the inner disk. This is opposite to the results presented here, where the inner disk starts to align with the plane of outer disk due to the excess of angular momentum of the infalling material.

The biggest idealization in our model is certainly the simplified initial condition of a two-phase medium of empty space and isolated gas cloudlet.   
As discussed in KGD20, a real Giant Molecular Cloud (GMC) is filamentary \citep{Andre2010}, and a star likely encounters assemblies of cold gas deviating from spherical symmetry. 
Nevertheless, observations of luminosity bursts for sources associated with filamentary arms such as FU Orionis or Z CMa \citep{Liu2016,Liu2018} strongly suggest the occurrence of late accretion events. 
Consistently, \cite{Pineda2020} reported the presence of an impressive $10 000 \unit{AU}$ gaseous streamer for the binary system Per-emb-2 (IRAS 03292+3039) with the IRAM NOrthern Extended Millimeter Array (NOEMA). 
The streamer was likely responsible for past accretion bursts as the chemical abundances are inconsistent with the relatively low present-day luminosity. 

Arguably, the best candidate for an infall-induced misaligned disk in action is SU Aur. \cite{Ginski2021} recently reported the presence of disk shadows together with extended filamentary arms that are strikingly similar to what is seen in our simulations. Moreover, \cite{Huang2021} showed the presence of extended arms around the star-disk system of GM Aur, which the authors interpret as a sign of infalling motions similar to the cases of AB Aur \citep{NakajimaGolimowski1995,Grady1999,Fukagawa2004} and SU Aur.

\subsection{Likelihood of late accretion events}
The idealized models investigated in this paper demonstrate that second-generation disks can form around an already existing star-disk system of a Herbig star during a late encounter event. 
Both computational models \citep{OffnerMcKee2011,Padoan2014,Jensen2018} as well as observations of accretion bursts \citep[e.g.][]{Kenyon1990,Evans2009} show that stars can undergo accretion events after its initial formation phase.   
Recently, \cite{Alves2020} reported on a possible candidate for simultaneous star and planet formation. [BHB2007] 1 hosts a Class II disks while also being connected to the larger scale environment through filamentary arms that strongly indicate ongoing infall of fresh gas and dust.
In particular, analyzing magnetohydrodynamical models that successfully reproduce the stellar initial mass function (IMF) \cite{Pelkonen2020} find that $\approx 90\%$ of the mass of intermediate-mass stars stems from gas unbound to the progenitor core. 
In contrast, low-mass stars tend to accrete relatively more mass during their initial formation phase from the corresponding prestellar core. 
Their models neglect stellar feedback and stellar radiation likely affects the formation process of higher mass stars \citep{Kuiper2015,Kuiper2016,Rosen2020}.
However, a higher probability of late accretion events for intermediate-mass stars than for low-mass stars could explain that tail- and arc- structures are more commonly observed around Herbig than T Tauri stars \citep[see discussion in][]{Dullemond2019}.

\subsection{Observations of misaligned configurations}
In this study, the infalling cloudlet interacts with a single star, but misaligned disks are also observed around binaries. 
A particularly interesting case is Oph IRS 43, which is a binary star with two circumstellar disks and a surrounding circumbinary disk. 
The two circumstellar disks are misaligned with respect to each other, as well as the surrounding circumbinary disk has an inclination of $\sim 70^{\circ}$ with respect to the binary orbit \citep{Brinch2016}.
Another prominent example of strong misalignment ($\sim 70^{\circ}$) between orbital plane of inner binary and outer disk is HD 142527 \citep{Marino2015}.  
Also, \cite{Alves2019} find a flattened envelope structure that appears to be misaligned to the orbital plane of the close binary [BHB2007] 11.
Carrying out star-forming models, \cite{Bate2018} finds misalignment of binary orbital plane and circumbinary disk in one case during an early stage of star formation for a short time period of $\sim 30$ kyr. 
Models similar to the ones presented here, but with infall on an existing binary system would help in understanding whether such misaligned configurations can be formed through late infall events.

\begin{figure}
    \includegraphics[width=\columnwidth]{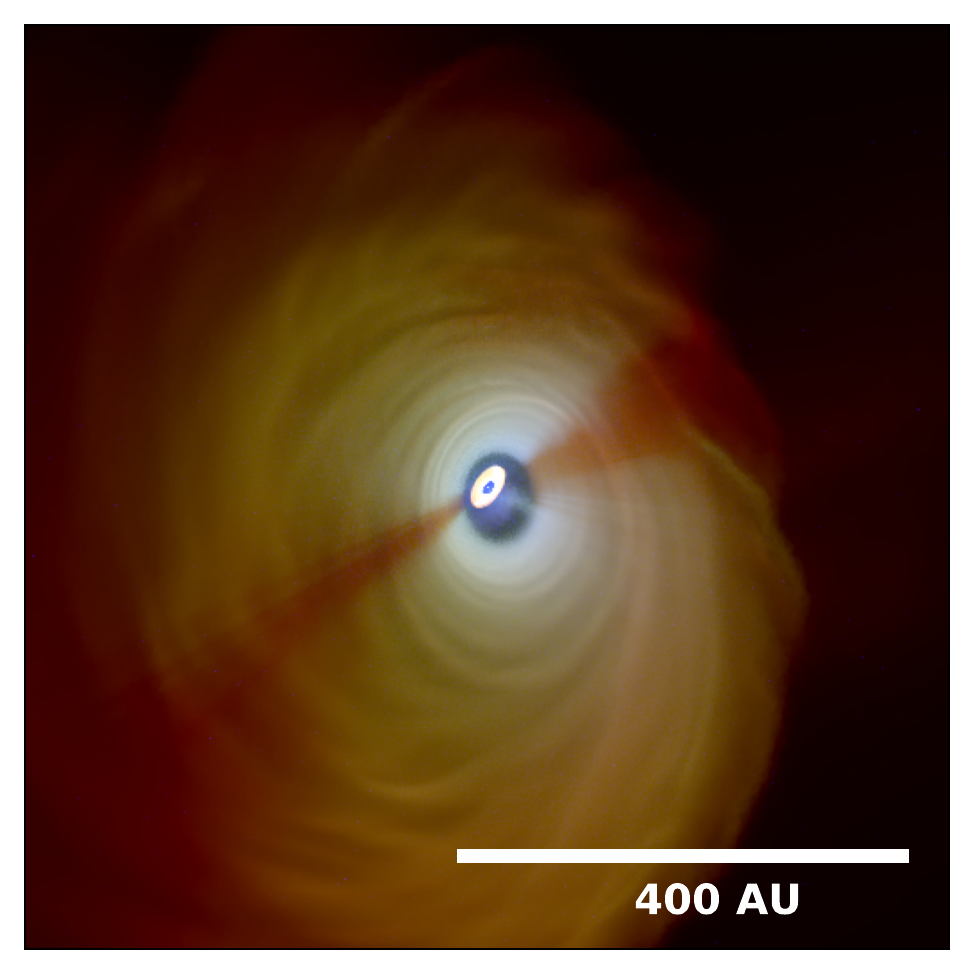} 
        \caption{Multi-wavelength RGB composite image of the run 3 for typical observational bands of $1.66\ \mu\mathrm{m}$ (blue, VLT H-band), $53.0\ \mu\mathrm{m}$ (green, SOPHIA/HAWC+ A-band), and $870.0\ \mu\mathrm{m}$ (red, ALMA band 7). The disks are observed from a distance of $140\ \mathrm{pc}$  under an angle of $45^\circ$ with respect to the main axis.}
        \label{fig:synthetic_shadow}
\end{figure}

In the last years, shadows located in the outer disk have been observed for a number of objects, such as HD 142527 \citep{Avenhaus2014}, DoAr 44 \citep{Casassus2018}, HD 135344B \citep{Stolker2016}, HD 100453 \citep{Benisty2017} or HD 143006 \citep{Benisty2018}. Moreover, \cite{Pontoppidan2020} reported on the variability of the great shadow in the Serpens Molecular Cloud \citep{PontoppidanDullemond2005} strongly hinting at a protoplanetary disk as the cause of the shadow.

In \Fig{synthetic_shadow} we present a multi-wavelength composite image for select bands of the observatories VLT ($1.66\ \mu\mathrm{m}$), SOPHIA/HAWC+ ($53.0\ \mu\mathrm{m}$), and ALMA ($870\ \mu\mathrm{m}$). 
To get this composite image, we post-process the results of our hydrodynamical runs with the Monte-Carlo (MC) code \polaris\footnote{\url{http://www1.astrophysik.uni-kiel.de/~ polaris/}}\ \citep{Reissl2016} that can handle the native Voronoi mesh of \arepo.
\polaris\ self-consistently calculates the dust temperature assuming that the dust component is in equilibrium with the radiation field.
For the radiation we utilize the central star and an additional diffuse interstellar radiation field (ISRF) \citep{Reissl2016,Reissl2020}. The star has an effective temperature of $T_{\mathrm{eff}}=9972\ \mathrm{K}$ and a radius of $R=1.0\ R_{\odot}$ leading to a luminosity of $L=8.2\ L_{\odot}$ while the ISRF follows the parametrization of \citep[][]{Mathis1983} with $G_0=1$ typical for the Milky Way. For the dust we assume spherical grains with a composition of silicates and graphite and apply the canonical MRN Milky Way dust with a minimal grain size of $a_{\mathrm{min}}=5\ \mathrm{nm}$ and a power-law size distribution $n(a)\propto a^{-3.5}$ \citep{Mathis1977}. 
However, circumstellar disks are environments that seem to support significant grain growth \citep{Ossenkopf1994,Kataoka2017}. 
Hence, we use an upper grain size of $a_{\mathrm{max}}=10\ \mu\mathrm{m}$ \citep[][]{Kataoka2017}. The dust mass to gas mass ratio is uniformly set to $1\ \%$ \citep{Bohlin1978} everywhere in the mesh.\\
Finally, we calculate intensity maps via a MC approach \citep{Reissl2016,Reissl2020} for scattered stellar radiation in the near-IR and via a ray-tracing scheme in the sub-mm for the thermal dust emission \citep[see Appendix B in][for details]{Reissl2019}. Here, we zoom into the inner $800\ \mathrm{AU}$ around the star and assume a distance to the observer of $140\ \mathrm{pc}$.

The synthetic observations of our models confirm that misalignment between an inner and outer disk can cast shadows on the outer disk as shown by \cite{Marino2015}. Similar to the  UV and mid-IR observations presented e.g. in \cite{Marino2015} and \cite{Benisty2017}  we note that the shadows of scattered light on the opposite sides of the outer disk are not exactly parallel but have a slight offset. We also emphasize that the gaseous stream from the outer disk onto the inner disk is clearly detectable in multiple bands of our synthetic image. In addition we see that the dust emission in the mid-IR and sub-mm is also decreased by a factor of about three in the shadowed regions, which means that not all of the light is blocked by the inner disk. 
We expect a larger shadow width in the outer disk for a more realistic treatment of the thermodynamics in the inner disk (see section 4.1) because a non-isothermal inner disk would be flared and more puffed up.  

Nevertheless even for the thin isothermal inner disks in our simulations, we find that the shielding of radiation by the inner disk has a significant impact on the dust temperature distribution of the outer disk. 
Considering the differences in eccentricities between inner and outer disk, we also expect that these differences are detectable in CO velocity channel maps. 
However, a detailed analysis of these effects goes beyond the scope of this paper. 
We analyze and discuss the observational constraints in more detail in a follow-up study.

For the observed systems, the inner disk is expected to be smaller than assumed in our models, likely of $\sim 1 \unit{AU}$ to $\sim 10 \unit{AU}$ in size.  The inner disk is tapered off at radii less than $\sim 20 \unit{AU}$ to avoid extensive computational costs. First, the hydrodynamical time step scales with radial distance in a Keplerian disk according $\Delta t \propto r^{1.5}$, which slows down the simulation when resolving the dynamics at small radii. Second, and more relevant in our case, for the assumed density scaling of $\Sigma \propto r^{-p}$ with $p=1.5$ or $p=1$ evolving smaller radii with the same resolution as used in this study requires modeling many more cells at the high densities in the inner part of the disk, and hence increases the computational costs.    
However, given that a second-generation disk with similar properties forms in scenarios without (see KGD20) and with the presence of an inner disk (this study), we expect that outer disks can also form around smaller disks than assumed in this work.  
The results of this study using a fixed cloudlet configuration and a constant impact parameter for all runs predominantly demonstrates that infall with substantial inclination angle can lead to the formation of misaligned systems. 
The size of the outer disk can be smaller for lower impact parameters as previously shown in KGD20.

Our results demonstrate that retrograde infall causes the inner disk to shrink over time. 
Due to conservation of angular momentum, shrinking is more efficient for lower inclination angles, but still occurs for substantial infall angles. 
In cases of retrograde infall with low angle, infall can hence lead to a configuration that is more of a misaligned ring-disk system. 
Recently, \cite{Kraus2020} reported on such a configuration around GW Orionis \citep[see also][]{Bi2020}. 
\cite{Kraus2020} attribute the misaligned configuration in GW Orionis to disk tearing \citep{Nixon2013} induced by the triple star system at the center. 
The crucial question is, however, whether breaking at a relatively large radius of $\approx 40 \unit{AU}$ can actually occur for low viscosities (i.e., low values of the Shakura-Sunyaev parameter $\alpha_{\rm SS}$ \citep{Shakura-Sunyaev1973}) that are expected in protoplanetary disks. 
The formula defined for the bending-wave regime, where $\alpha_{\rm SS}<H/r$ suggests a significantly smaller breaking radius than the formula for the diffusive regime that was used to estimate the breaking radius (see equation 10 and A3 in \cite{Nixon2013}, see also equations 2 and 3 in \cite{Facchini2018}).
\cite{Kraus2020} assume relatively large values of $\alpha_{\rm SS}\approx 0.01$ to $0.02$ for their simulations and apply the formula defined for the diffusive regime, where $\alpha_{\rm SS}$ exceeds the aspect ratio of scale height to radius ($\alpha_{\rm SS}>H/r$) for their analytical considerations of the disk breaking radius (equation A3 in \cite{Nixon2013}, see also equation 2 in \cite{Facchini2018}). 
In fact, \cite{Smallwood2021} carried out Smoothed Particle Hydrodynamics (SPH) simulations with varying viscosities in the disk and found no breaking in the bending-wave regime. 
They suggest instead that the stellar multiple leads to a warped inner disk after a planet carved a gap in the disk.

Our results strongly suggest that configurations of misaligned inner and outer disks can also arise after infall events.
Especially, GW Orionis is located in a clustered environment of Lambda Orionis, where the probability of infall or gas-encounter events is expected to be higher than for isolated objects.

Similar to the results found by \cite{Wijnen2017a}, the inner disk tends to align with the orientation of the outer disk for misaligned infall with sufficient angular momentum, eventually erasing the initial misalignment between the disks.
However, the inner disk can only align with the outer disk if angular momentum can be transferred between the disks. 
Considering an early onset of planet formation \citep{Harsono2018}, as well as the frequent observation of ring structures in young disks \citep{ALMAHLTau2015,Andrews2018} the presence of a deep gap carved by an accreting planet might quench alignment of the disk inside the radius of the planet.
Note that different from the scenario of planet-induced misalignment considered by \cite{Zhu2019}, the orientation of the inner disk would remain unchanged.

\subsection{Future goals for modeling}
Ultimately, to model the infall event in a more realistic setup in the future requires models that consistently solve the small-scale physical processes on AU-scales, while simultaneously account for the protostellar environment. 
Multi-scale simulations similar to the ones presented in \citet{Kuffmeier2017,Kuffmeier2019}, but for later times of evolution are therefore the crucial next step.
Carrying out simulations for a sample of objects also allows to obtain statistics on the occurrence rate of encounter events, as well as to constrain the role of infall in possibly warping the disk around binary stars, thus causing misalignment of the binary orbit and disk.

\section{Conclusion}
Hydrodynamical models with the moving-mesh code \arepo\ demonstrate that for a star with an already existing disk, infall of a gaseous cloudlet induces the formation of a second-generation disk around the inner primordial disk. 
In the likely event of new material approaching the inner disk on an inclined trajectory with respect to the midplane of the inner disk, the orientation of the new outer disk is misaligned to the inner disk. 
As a consequence of the misaligned configuration, the inner disk casts a shadow on the outer disk as shown in synthetic observations, which is consistent with observations of several transition disks.  
The orientation of the outer disk is governed by the inclination angle of the infalling material, and hence the misalignment between inner and outer disk can be substantial.

In the case of retrograde infall the outer disk counter-rotates with respect to the inner disk and carves a larger gap between the two disks than for prograde infall. 
The resulting gap can become larger than the size of the inner disk, which is consistent with observations of several transition disks.
For low inclination angles, retrograde infall removes angular momentum from the inner disks that shrinks the inner disk and enhances accretion in comparison to prograde infall.

As a by-product of infall, the forming outer disks are associated with characteristic arc- and spiral structures on scales of several 100 to 1000 AU. 
The outer disk forms with high eccentricity of $0.2$ to $0.4$ mostly independent of the inclination angle of infall. 
By the end of the simulations at $t=150 \unit{kyr}$, the eccentricity of the outer disk is $0.3$ to $0.4$ and the disk's outer edge becomes more sharp independent of the infall angle.     
We therefore predict that infall-induced disks have significant eccentricities, which are expected to be detectable in CO channel maps.

In the inner disk, infall triggers modest eccentricity of $e=0.05$ to $\approx 0.1$ in all cases. 
Smaller inclination angles of infall cause larger and more variable eccentricity than infall  with high inclination.

During the evolution, the inner disk tends to align with the orientation of the outer disk due the higher angular momentum of the infall material on the larger-scale disk. 
We show that alignment with the outer disk occurs faster for a lower initial mass of the inner disk, consistent with angular momentum conservation.

Furthermore, infall triggers bumps in the density profile of the inner disk, which could possibly support trapping of dust grains, and hence provide favorable conditions for dust growth and ultimately planet formation.  
Finally, if misaligned disks are predominantly a consequence of infall we expect a higher fraction of misaligned disks and arc-structures around Herbig stars than T Tauri stars as recent star formation models indicate a higher probability of late infall for intermediate-mass than low-mass stars.

\begin{acknowledgements}
We thank the referee for a thorough review of an earlier version of the manuscript. In particular, we thank the referee for providing constructive feedback and useful suggestions that improved the quality of this work.
We thank Sebastian Wolf for useful comments on the manuscript especially on the observational aspects. 
This research project of MK is mainly supported by a research grant of the Independent Research Foundation Denmark (IRFD) (international postdoctoral fellow, project number: 8028-00025B). 
MK acknowledges the support of the DFG Research Unit `Transition Disks' (FOR 2634/1, DU 414/23-1). MK also gratefully acknowledges that this project has received funding from the European Union's Framework Programme for Research and Innovation Horizon 2020 (2014-2020) under the Marie Sk{\l}odowska-Curie Grant Agreement No.\, 897524. FG acknowledges support from DFG Schwerpunktprogramm SPP 1992 `Diversity of Extrasolar Planets' grant DU 414/21-1. SR acknowledges funding from the Deutsche Forschungsgemeinschaft (DFG, German Research Foundation) -- Project-ID 138713538 -- SFB 881 ``The Milky Way System'' (sub-projects B1, B2, and B8)), from the Priority Program SPP 1573 ``Physics of the Interstellar  Medium''  (grant  numbers  KL  1358/18.1,  KL  1358/19.2), and acknowledges also support from the DFG via the Heidelberg Cluster of Excellence {\em STRUCTURES} in the framework of Germany’s Excellence Strategy (grant EXC-2181/1 - 390900948).
\end{acknowledgements}

\bibliography{general}
\bibliographystyle{aa}

\appendix

\section{Evolution sequences of run 1, 3, 4, 10 and 11}
\begin{figure*}
    \includegraphics[width=\textwidth]{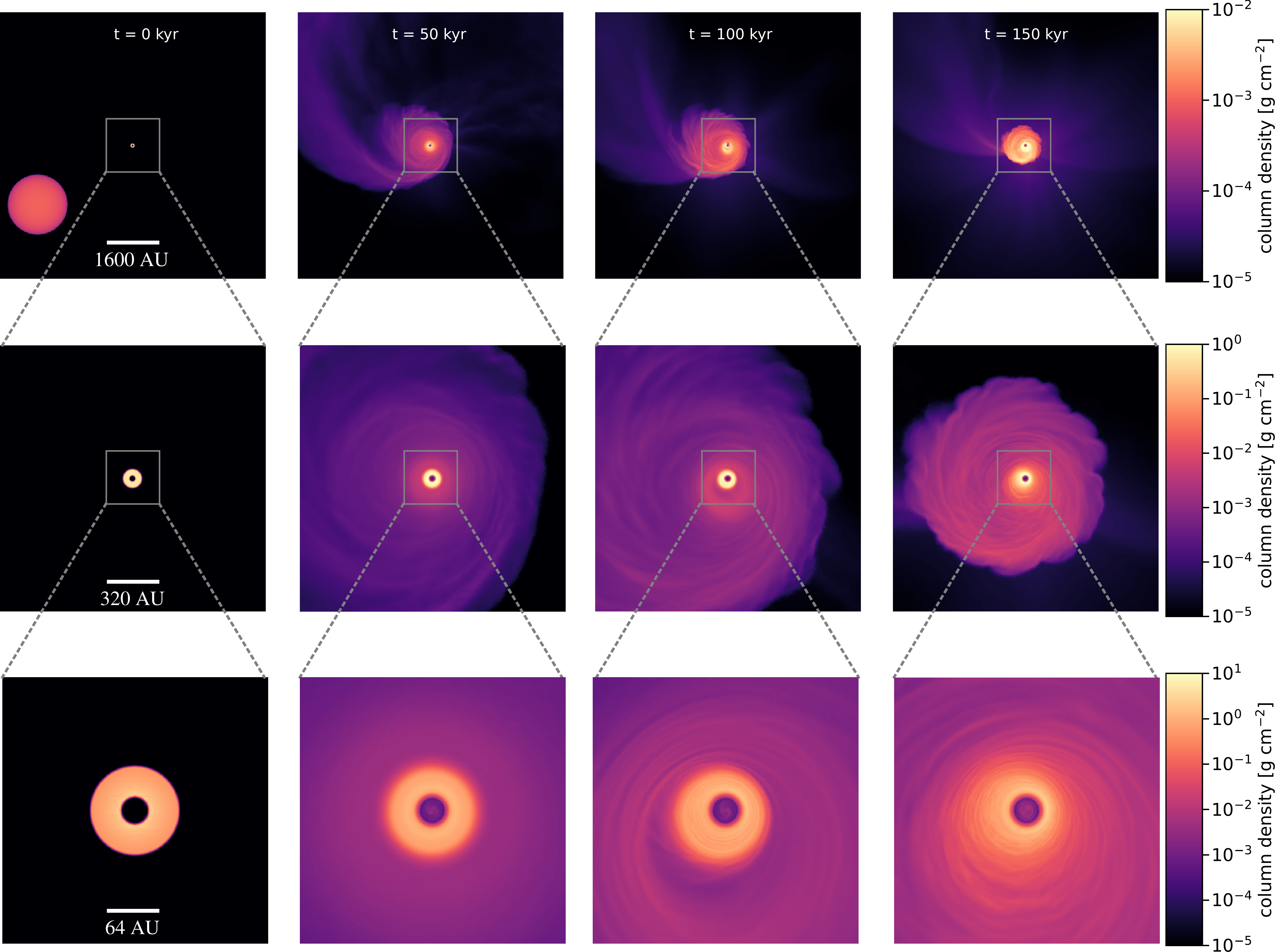} 
        \caption{Same as \Fig{1p5_seq_000}, but for run 1.}
        \label{fig:1p5_seq_align}
\end{figure*}

\begin{figure*}
    \includegraphics[width=\textwidth]{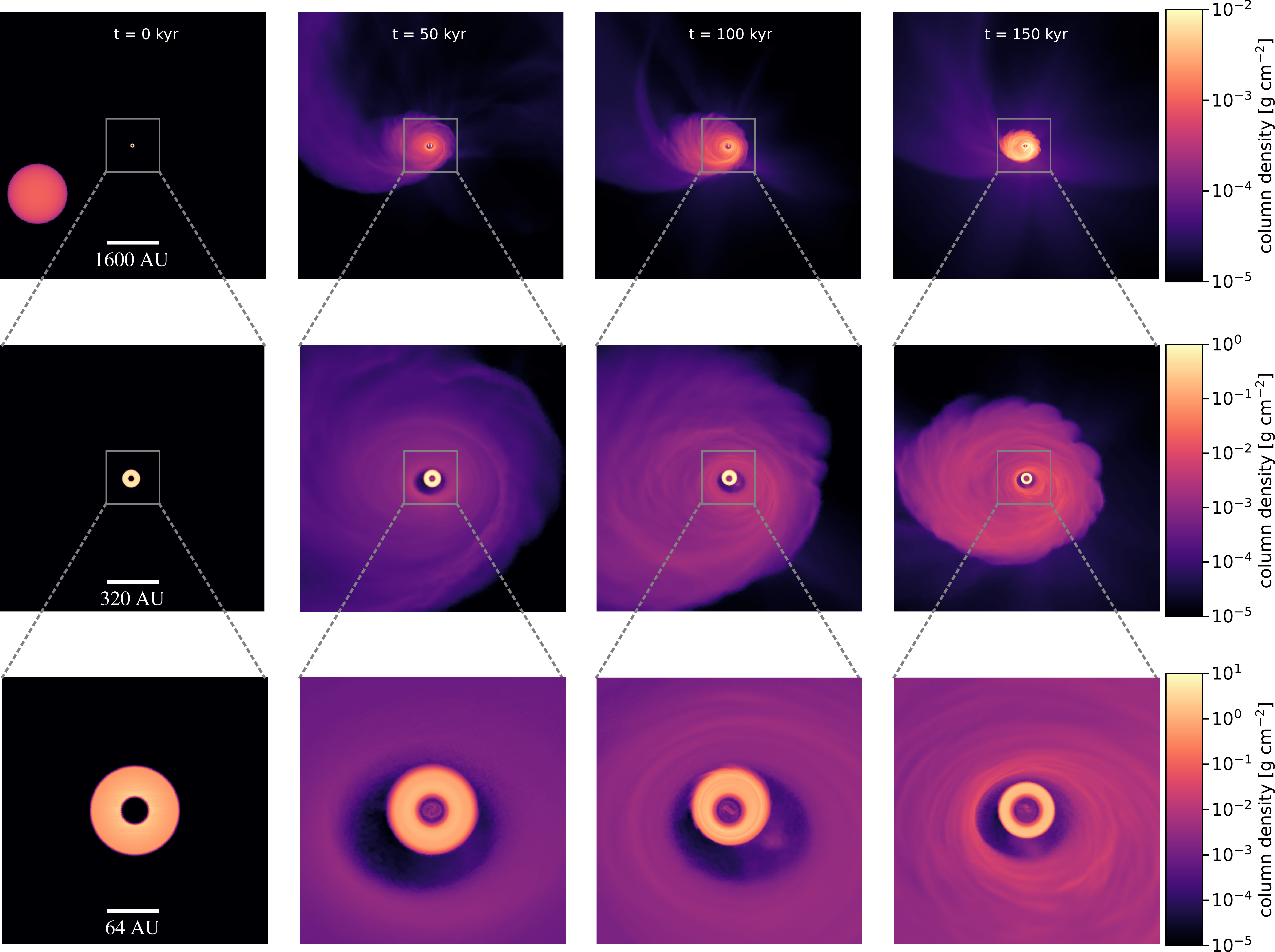} 
        \caption{Same as \Fig{1p5_seq_000}, but for run 3.}
        \label{fig:1p5_seq_035}
\end{figure*}

\begin{figure*}
    \includegraphics[width=\textwidth]{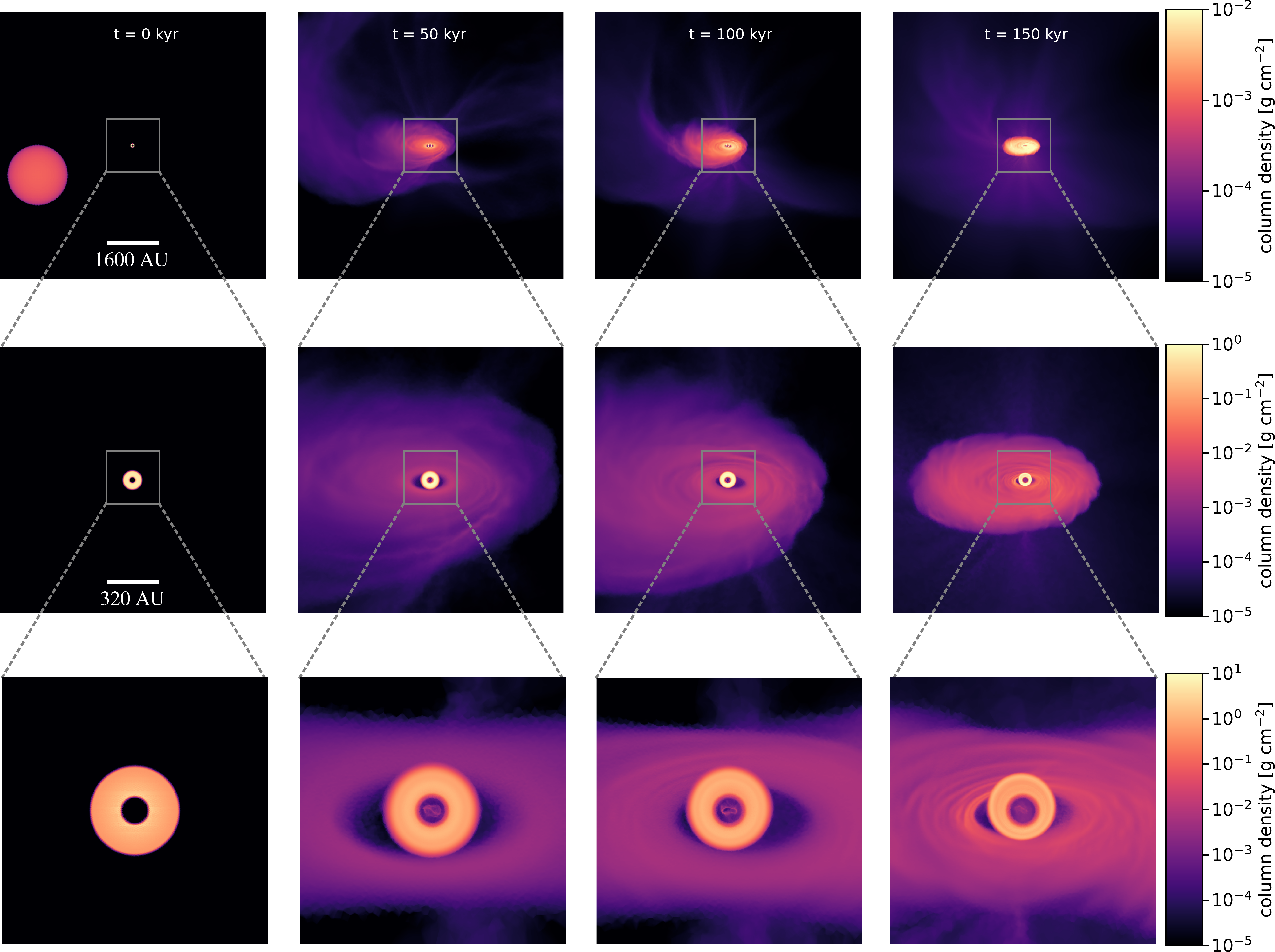} 
        \caption{Same as \Fig{1p5_seq_000}, but for run 4.}
        \label{fig:1p5_seq_060}
\end{figure*}

\begin{figure*}
    \includegraphics[width=\textwidth]{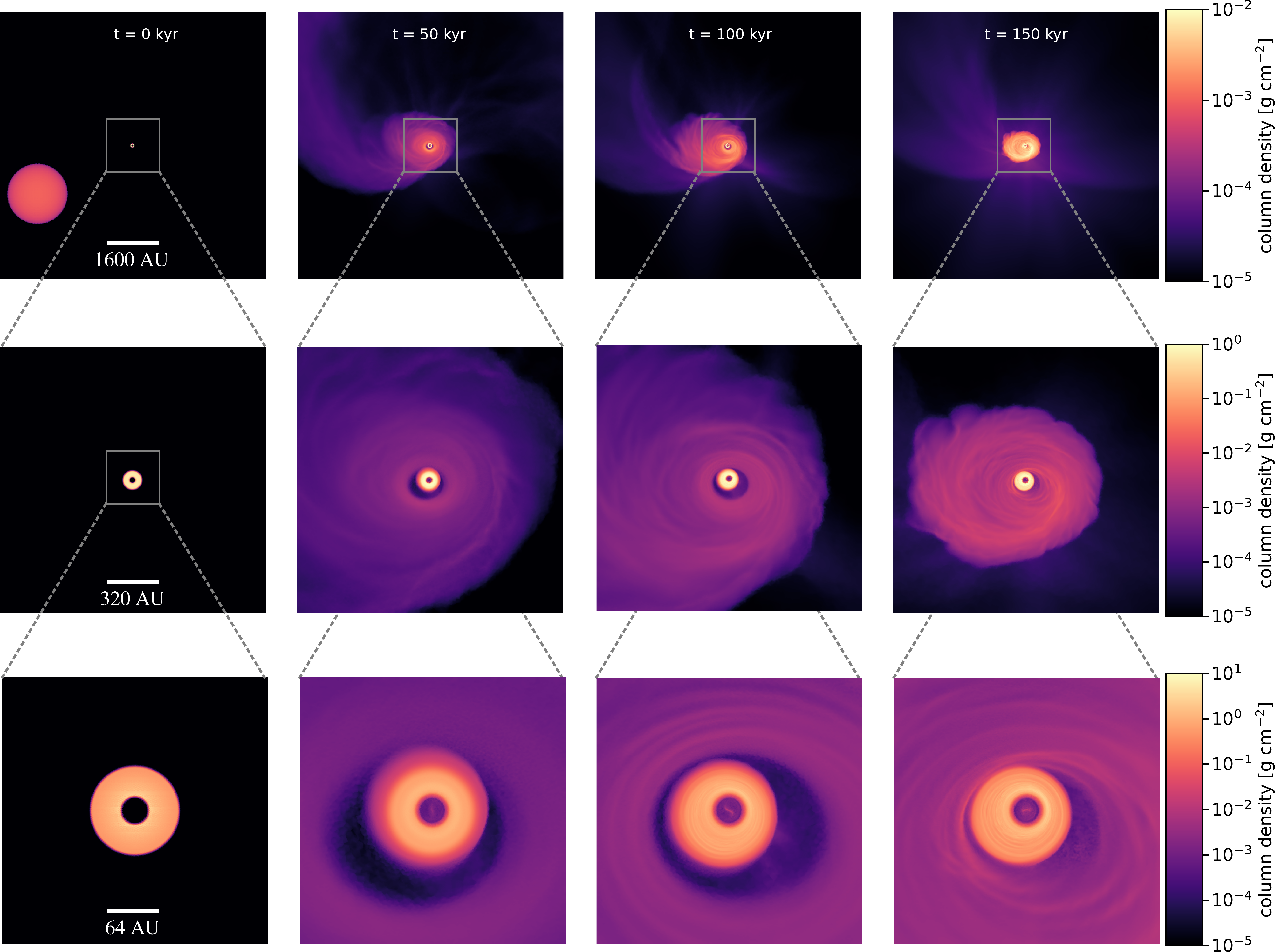} 
        \caption{Same as \Fig{1p5_seq_000}, but for run 10.}
        \label{fig:1p5_seq_035a}
\end{figure*}

\begin{figure*}
    \includegraphics[width=\textwidth]{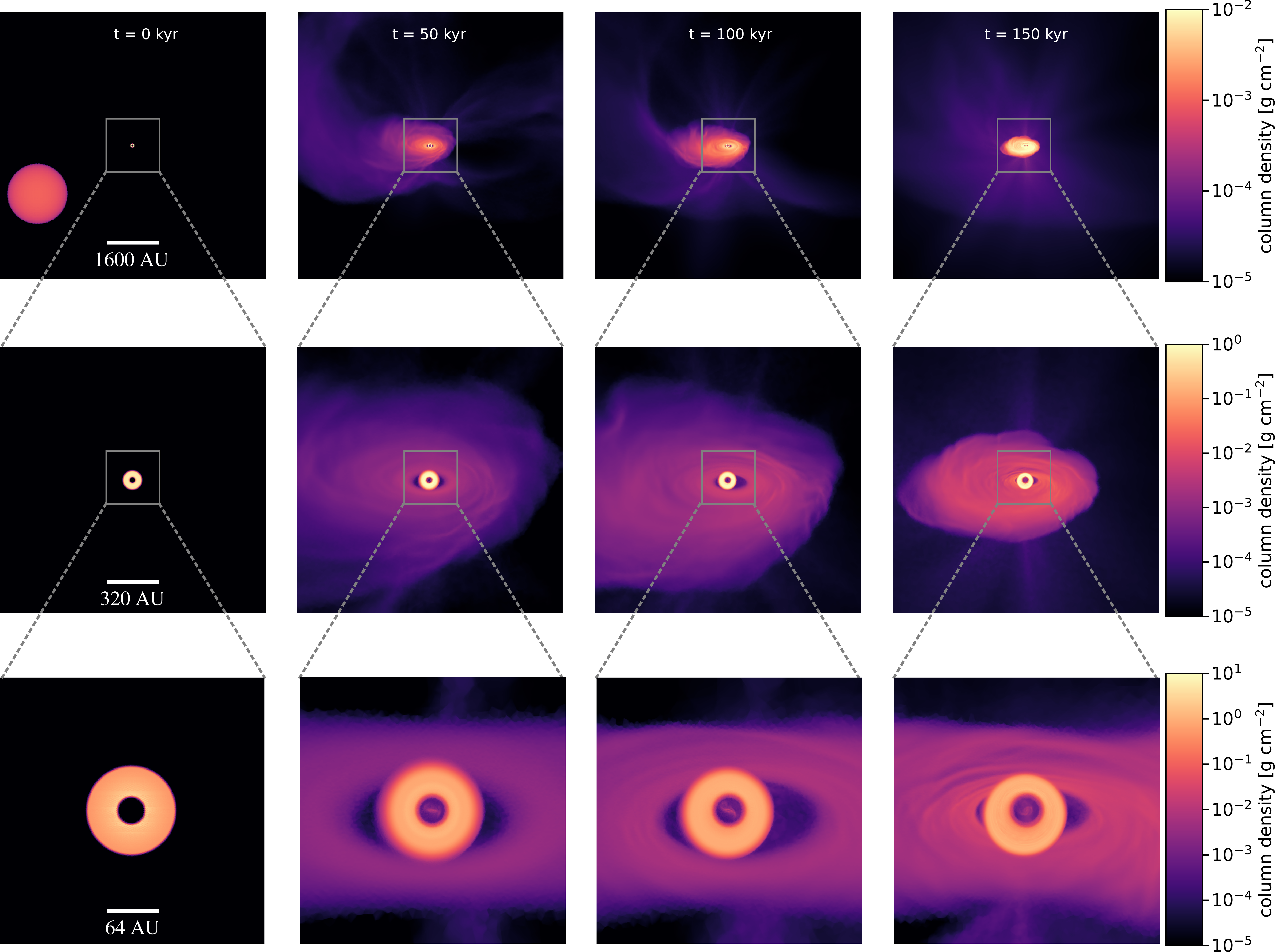} 
        \caption{Same as \Fig{1p5_seq_000}, but for run 11.}
        \label{fig:1p5_seq_060a}
\end{figure*}

Here we show the evolution of the column density for the run with infalling angle of $\alpha=0^{\circ}$, but prograde infall (run 1; \Fig{1p5_seq_align}). Moreover, we show the sequence of retrograde infall for $\alpha=35^{\circ}$ (run 3; \Fig{1p5_seq_035}) and $\alpha=60^{\circ}$ (run 4; \Fig{1p5_seq_060}). 
\Figure{1p5_seq_align} illustrates that the inner disk remains large with respect to the retrograde case in run 2 (\Fig{1p5_seq_000}). 
Also, there is no long-lasting gap forming between inner and outer disk in the prograde case as discussed in the main text.
The sequence for infalling angles $\alpha=35^{\circ}$ and $\alpha=60^{\circ}$ of both retro- and prograde infall is illustrated in \Fig{1p5_seq_035}, \Fig{1p5_seq_060}, \Fig{1p5_seq_035a} and \Fig{1p5_seq_060a}. The sequences illustrate that the outer disk forms misaligned with respect to the orientation of the primordial inner disk and that the inner disk becomes smaller for retrograde infall. 

\end{document}